\documentclass[useAMS,usenatbib,letterpaper]{mn2e}
\usepackage[totalwidth=480pt,totalheight=680pt,layoutvoffset=0.5cm]{geometry}
\usepackage{graphicx}
\usepackage{amsmath}
\usepackage{astrobib_mnras2e}
\usepackage{times}
\usepackage{fixltx2e}

\newcommand{\C}{{\bf C}}

\newcommand{\cS}{{\bf S}}
\newcommand{\iC}{{\bf \Psi}}
\newcommand{\iCh}{\hat{\bf \Psi}}
\newcommand{\x}{{\bf x}}

\newcommand{\he}{{\hat{e}}}
\newcommand{\hpsi}{\hat{\psi}}
\newcommand{\normal}[2]{{\cal N}(#1, #2)}

\DeclareMathOperator*{\argmax}{argmax}

\title{Estimating sparse precision matrices}
\author[Padmanabhan et al]{ 
  \parbox{\textwidth}{
Nikhil Padmanabhan$^1$, 
Martin White$^2$,
Harrison H. Zhou$^3$,
Ross O'Connell$^4$}\vspace{0.4cm}\\
\parbox{\textwidth}{
\scriptsize $^1$ Dept. of Physics, Yale University, New Haven, CT 06511 \\
\scriptsize $^2$ Dept. of Physics and Astronomy, U.C. Berkeley, Berkeley, CA, 94720 \\
\scriptsize $^3$ Dept. of Statistics, Yale University, New Haven, CT 06511 \\
\scriptsize $^4$ McWilliams Center for Cosmology, Carnegie Mellon University, 5000 Forbes Ave, Pittsburgh, PA 15213, USA \\
}
}

\begin{document}

\maketitle
\begin{abstract}
We apply a method recently introduced to the statistical literature to directly
estimate the precision matrix from an ensemble of samples drawn from a 
corresponding Gaussian distribution. Motivated by the observation that cosmological 
precision matrices are often approximately sparse, the method allows one to exploit this
sparsity of the precision matrix to more quickly converge to an asymptotic
$1/\sqrt{N_{\rm sim}}$ rate while simultaneously providing an error model for all of the
terms.
Such an estimate can be used as the starting point for further regularization
efforts which can improve upon the $1/\sqrt{N_{\rm sim}}$ limit above, and incorporating
such additional steps is straightforward within this framework.
We demonstrate the technique with toy models and with an example motivated by
large-scale structure two-point analysis, showing significant improvements in
the rate of convergence.
For the large-scale structure example we find errors on the precision matrix
which are factors of 5 smaller than for the sample precision matrix for
thousands of simulations or, alternatively, convergence to the same error level
with more than an order of magnitude fewer simulations.
\end{abstract}
\begin{keywords}
cosmological parameters --- large-scale structure of Universe
\end{keywords}

\section{Introduction}
\label{sec:intro}

Frequently in astrophysics and cosmology the final step in any analysis
is to compare some summary of the data to the predictions of a theoretical
model (or models).  In order to make this comparison a model for the
statistical errors in the compressed form of the data is required.  The
most common assumption is that the errors are Gaussian distributed, and
the covariance matrix, $C_{ij}$, is supplied along with the data.
In order for precise and meaningful comparison of theory and observation,
the theoretical model, summary statistics derived from the data and $C_{ij}$
must all be accurately determined.  A great deal of work has gone into all
three areas.

There are three broad methods for determining such a covariance matrix.
First, there may be an analytical or theoretical description of $C_{ij}$
that can be used \cite[see][for some recent examples and further
references]{Grieb15, OConnell15, Pearson15}.  
Secondly, we may attempt to infer $C_{ij}$ from the
data itself.  This is often referred to as an `internal error estimate'
and techniques such as jackknife \citep{Tuk58} or bootstrap \citep{Efr79}
are commonly used.
Thirdly, we may attempt to infer $C_{ij}$ by Monte-Carlo simulation.  This
is often referred to as an external error estimate.
Combinations of these methods are also used.

Our focus will be on the third case, where the covariance matrix is
determined via Monte-Carlo simulation.  The major limitation of such
methods is that the accuracy of the covariance matrix is limited by the
number of Monte-Carlo samples that are used with the error typically
scaling at $N_{\rm sim}^{-1/2}$
\citep[see, e.g.][for recent discussions in the cosmology context]{DodSch13,
Tay13,TayJoa14}.
These errors manifest themselves both as `noise' and `bias' in the
sample covariance matrix and the inverse covariance matrix (which we shall
refer to as the `precision matrix' from now on and denote by $\iC$).

We consider the specific case where the precision matrix is sparse, 
either exactly or approximately. This may happen even when the covariance
matrix is dense, and occurs generically when the correlations in the 
covariance matrix decay as a power law (or faster). 
It is worth
emphasizing that any likelihood analysis requires the precision 
matrix, and not the covariance matrix. 
We present an
algorithm that can exploit this sparsity structure of the precision
matrix with relatively small numbers of simulations. 

The outline of the paper is as follows.
In Section \ref{sec:notation} we introduce our notation and set up the
problem we wish to address.
Section \ref{sec:method} introduces the statistical method for entry-wise
estimation of the precision matrix, plus a series of refinements to the
method which improve the convergence.
Section \ref{sec:numerics} presents several numerical experiments which
emphasize the issues and behavior of the algorithm, both for a toy model
which illustrates the principles behind the algorithm and for
a model based on galaxy large-scale structure analyses.
We finish in Section \ref{sec:discussion} with a discussion of our results,
the implications for analysis, and directions for future investigation.

\section{Nomenclature and Preliminaries}
\label{sec:notation}

We will consider $p$-dimensional observations, $\x$, with a covariance
matrix $\C$ and its inverse, the precision matrix $\iC$; we differ from
the usual statistical literature where the covariance and precision matrices
are represented by $\mathbf{\Sigma}$ and $\mathbf{\Omega}$ in order to match
the usage in cosmology more closely.
Since we will need to consider both estimates of these matrices as well as
their (unknown) true value, we denote estimates with hats. Individual 
elements of the matrix are represented by the corresponding lowercase
Greek letters, e.g. $\psi_{ij}$. We will also want to consider normalized
elements of the precision matrix - we will abuse notation here and 
define these by $r_{ij} \equiv \psi_{ij}/\sqrt{\psi_{ii} \psi_{jj}}$.

\begin{table}
  \begin{tabular}{ll}
    \hline
    Notation & Description \\
    \hline
    $\normal{\mu}{\C}$ & Normal distribution with mean $\mu$, covariance $C$ \\
    $x \sim {\cal D}$ & $x$ distributed as ${\cal D}$ \\
    $\Vert \cdot \Vert_{F}$ & Frobenius matrix norm (see Eq.~\ref{eq:frobdef}) \\
    $\Vert \cdot \Vert_{2}$ & Spectral matrix norm (see after Eq.~\ref{eq:frobdef}) \\
    ${\bf A}^t$ & Transpose of ${\bf A}$ \\
    \hline
  \end{tabular}
  \caption{Summary of notation used in this paper}
  \label{tab:notation}
\end{table}

We denote the normal distribution with mean $\mu$ and covariance $\C$ by 
$\normal{\mu}{\C}$. The notation $x \sim {\cal D}$ denotes a random 
variable $x$ with a probability distribution ${\cal D}$.
The Frobenius norm of a matrix is defined by 
\begin{equation}
  \Vert {\bf A} \Vert_F \equiv \Bigg( \sum_i \sum_j a_{ij}^2 \Bigg)^{1/2} =  \Bigg( {\rm Tr} A A^{t} \Bigg)^{1/2} \,
  \label{eq:frobdef}
\end{equation}
while the spectral norm, denoted by $\Vert \cdot \Vert_2$, is the largest singular value of the 
matrix. 
Table \ref{tab:notation} summarizes our notation.

The problem we consider is estimating the $p \times p$ precision matrix, $\iC$,
from $d$ independent samples\footnote{We use $d$ instead of $N_{\rm sim}$ in what follows for 
brevity.}, $\x_i$ where $1 \le i \le d$ and where $\x_i$
is a $p$ dimensional vector assumed to be drawn from a Gaussian distribution.
The usual approach to this problem has been to compute the sample 
covariance matrix
\begin{align}
  \cS = \frac{1}{d-1} \sum_{i=1}^{d} (\Delta \x_i) (\Delta \x_i)^t
\end{align}
where the superscript $t$ is the transpose, and $\Delta \x_i \equiv \x_i - (1/d) \sum_{i=1}^{d} \x_i$
is the difference vector. An unbiased estimate of the precision matrix is 
then 
\begin{align}
  \iCh = \frac{d-p-2}{d-1} \cS^{-1}
\end{align}
where the prefactor accounts for the mean of the inverse-Wishart distribution
(for a first application in cosmology, see
\citealt{HarSimSch07}).

\section{The Method}
\label{sec:method}

Our approach in this paper uses the observation that precision matrices in cosmology
are often very structured and sparse. Unfortunately, this structure is hard to exploit
if computing the precision matrix involves the intermediate step of computing the
covariance matrix. Our approach uses a technique, pointed 
out in \cite{Ren2015}, to directly compute the precision 
matrix from an ensemble of simulations. Unlike that work, which was interested in 
an arbitrary sparsity pattern, the structure of cosmological precision matrices is 
expected to be more regular, significantly simplifying the problem.

The steps in our algorithm are :
\begin{enumerate}
  \item Estimate the elements of the precision matrix entrywise.
  \item Smooth the precision matrix.
  \item Ensure positive-definiteness of the resulting precision matrix.
\end{enumerate}
Each of these are discussed in detail below.
It is worth emphasizing that 
the key insight here is the first step, and it is easy to imagine variants of the
remaining steps that
build off an entrywise estimate of the precision matrix.

\subsection{Entrywise estimates}

Consider a random vector $\x_i = (Z_1, Z_2,...,Z_p)$ drawn from a multivariate normal distribution 
with mean 0 (this is trivially generalized) and 
covariance matrix $\C$. Imagine partitioning the components into two sets $A$ and $A^c \equiv 
\{1,\ldots,p\}\setminus A$ with $Z_{A}$ denoting the subset of components in set
$A$. Consider the probability of $Z_{A}$ conditioned
on $Z_{A^c}$ (see Appendix \ref{app:la} for some useful identities)
\begin{equation}
  P(Z_{A} | Z_{A^c}) = \normal{- \iC_{A,A}^{-1} \iC_{A,A^c} Z_{A^c}}{\iC_{A,A}^{-1}} \,,
\label{eq:condprob}
\end{equation}
where $\iC_{A,B}$ represents the submatrix indexed by the sets of indices $A$ and $B$. This equation can 
be interpreted as linearly regressing $Z_A$ on $Z_{A^c}$ :
\begin{equation}
  Z_A = \beta Z_{A^c} + e_A
\label{eq:regress}
\end{equation}
where $\beta = -\iC_{A,A}^{-1} \iC_{A,A^c}$ and $\langle e_A e_A^t \rangle =
\iC_{A,A}^{-1}$. This interpretation is key to the algorithm presented here :
the inverse of covariance of the residuals of the above linear regression is
an estimate of a subset of the full precision matrix. 

The above equation also demonstrates how to make use of the sparsity of the
precision matrix. Note that the submatrix $\iC_{A,A^c}$ (that appears in
$\beta$) inherits the sparsity of the precision matrix and therefore, one only
need regress on a small number of elements of $Z_{A^c}$. To illustrate this
with a concrete example, consider a tridiagonal $\iC$ and $A = \{1\}$.
Eqs.~\ref{eq:condprob} and ~\ref{eq:regress} imply that $Z_1 = \beta Z_2 + e$
where, in this case, $\beta$ and $e$ are simply scalars, and $\langle e^2
\rangle = \psi_{1,1}^{-1}$.  
Given measurements
$(Z_1^{(1)},Z_2^{(1)},\cdots)$,
$(Z_1^{(2)},Z_2^{(2)},\cdots)$, $\ldots$,
$(Z_1^{(d)},Z_2^{(d)},\cdots)$,
we do an ordinary least-squares fit for $\beta$ estimating the error
$e^2$ and use $\psi_{1,1}=e^{-2}$.
The linear regression in this case requires $d
\gg 2$ observations to robustly determine $\beta$ and $e$, compared with $d \gg
p$ observations where $p$ is the rank of the precision matrix; this can
translate into a significant reduction in the required number of simulations.

We can now write down the first step of our algorithm. For every pair $1 \le i
< j \le p$, linearly regress $Z_{\{ij\}}$ on $Z_{\{k,l,\ldots\}}$ where the
${k,l,\ldots}$ are determined by the sparsity pattern of the matrix. For the
cases considered here, we perform an ordinary least squares regression, which
guarantees that $\hat{\beta}$ and $\hat{e}$ are independent \citep{greene2003}.  From the
residuals, we form the $2 \times 2$ covariance matrix, which can be inverted to
get an estimate of the precision matrix elements $\psi_{ii}$, $\psi_{ij}$
and $\psi_{jj}$ :
\begin{equation}
\begin{pmatrix}
  \hpsi_{ii} & \hpsi_{ij} \\
  \hpsi_{ji} & \hpsi_{jj} \\
\end{pmatrix}^{-1}
= \frac{1}{d-K} \sum
\begin{pmatrix}
  \he_{i}^2 & \he_{i} \he_{j} \\
  \he_{i} \he_{j} & \he_{j}^2 
\end{pmatrix}
\end{equation}
where the sum is over the $d$ observations (the index is suppressed
to avoid confusion) and $K$ is the number of variables
being regressed on. 
Note that this gives us estimates of $\psi_{ii}$, $\psi_{ij}$
and $\psi_{jj}$. 
While it is possible to directly estimate $\psi_{ij}$, we have 
found it more robust to estimate 
$r_{ij} = \psi_{ij}/\sqrt{\psi_{ii} \psi_{jj}}$. This combination reduces the
finite-sample corrections\footnote{For instance, the $d-K$ factor cancels in
its computation} and as we demonstrate in the next section, achieves its
asymptotic distribution $(r - \hat{r}) \sim \normal{0}{\sqrt{1-r^2}/\sqrt{d}}$
for relatively small values of $d$. We therefore use these pairwise regressions
to only compute $r_{ij}$. 

In order to compute the $\psi_{ii}$, we repeat the above regression analysis
for $A={i}$. Note that we could use the values of $\psi_{ii}$ calculated in
the previous step, but working in the single variable case simplifies the
analysis of the properties of the estimator. Defining 
\begin{equation}
  s^2 = \frac{1}{d-K} \sum {\he_i^2}
\end{equation}
one can show \citep{greene2003} that 
\begin{equation}
  (d-K) s^2 \psi_{ii} \sim \chi^2_{d-K}
\end{equation}
and therefore the estimator
\begin{equation}
  \hat{\psi}_{ii} = \frac{d-K-2}{\sum \he_i^2}
\end{equation}
is distributed as a scaled inverse $\chi^2$ distribution with $d-K$
degrees of freedom.

At the end of this first step, our estimate of the precision matrix can
be written as 
\begin{equation}
  \iCh = {\cal D} {\cal R}_0 {\cal D}
\end{equation}
where ${\cal D}$ is a diagonal matrix with
${\cal D}_{ii}=\sqrt{\hat{\psi}_{ii}}$
while ${\cal R}_0$ has $r_{ij}$ on the off-diagonals and $1$ on the diagonal.

\subsection{Smoothing}

We expect the covariance and precision matrices we encounter in cosmology
to be relatively ``smooth'', which we can therefore hope to use to reduce
the noise in the entrywise estimates.
Such a smoothing operation can take many forms. If we could construct a
model for the covariance/precision matrix, one could directly fit the
entrywise estimates to the model.
For example, on large scales, one might use a Gaussian model with free
shot noise parameters \citep[]{Xu2012,OConnell15}.
In the absence of a model, one might use a non-parametric algorithm to smooth
the precision matrix, respecting the structure of the matrix.
For the examples we consider in this paper, we use a cubic smoothing spline
and smooth along the off-diagonals of the matrix, with the degree of
smoothing automatically determined by the data using a cross-validation
technique.
Since we believe that such a smoothing procedure may be generically useful,
Appendix~\ref{app:cubic} contains an explicit description of the algorithm.

\subsection{Maximum-Likelihood Refinement}

The estimate of the precision matrix is ``close'' to the true precision matrix in a
Frobenius sense. However, since the matrix was estimated entrywise, there is no
guarantee that ${\cal R}_0$ (and therefore $\iCh$) is positive definite\footnote{The
diagonal matrices ${\cal D}$ are positive definite by construction.}. Our goal in 
this section is to find a refined estimate ${\cal R}$ that is both positive definite 
and close to ${\cal R}_0$.

A natural approach is to choose ${\cal R}$ to maximize its posterior 
$P({\cal R} | \cS)$ given an observed sample covariance matrix and the prior on ${\cal R}$
from ${\cal R}_0$. Again assuming Gaussianity, the likelihood of $\cS$ (ignoring irrelevant constants) is 
\begin{equation}
  2 \log P(\cS | \iC)  = d \big( \log \det \iC  - {\rm tr}\, \cS \iC \big) \,
\end{equation}
while we take the prior on ${\cal R}$ to be 
\begin{equation}
  2 \log P(\iCh) = -d \Vert {\cal R} - {\cal R}_0 \Vert^2_F
\end{equation}
where we assume that the error on $r_{ij}$ is $d^{-1/2}$. We ignore the $r$ dependence
on the error to avoid potentially biasing these results with noise from the estimate; 
note that this error estimate is a conservative estimate. For similar reasons, we hold the
diagonal matrix ${\cal D}$ fixed. Putting this together, our maximization problem can 
be written as 
\begin{equation}
{\cal R} = \argmax_{\substack{
    r_{ij}, (ij) \in {\cal J} \\
    {\cal R} \succ 0 
  }}
  \Bigg[ \log \det {\cal R} - {\rm tr} \big({\cal D} \cS {\cal D} {\cal R}\big)- 
    \Vert {\cal R} - {\cal R}_0 \Vert^2_F
  \Bigg]
\end{equation}
where ${\cal J}$ is the set of indices of the nonzero elements of ${\cal R}$ (as determined
by the sparsity of the matrix) and ${\cal R} \succ 0$ represents the space of positive 
definite matrices. We also observe that while we have 
fixed the relative weights of the likelihood and prior terms here based
on the expected error on $r$, it is possible in principle
to weight these contributions differently. We leave this possibility 
open for future applications. 
We also note that this refinement step is reminiscent of the {\it Scout} estimator of \cite{Witten2009}.

We perform this optimization using standard techniques; details of the algorithm are
in Appendix~\ref{app:mlrefine}.

\section{Numerical Experiments}
\label{sec:numerics}

\subsection{Model Covariance and Precision Matrices}

\begin{figure}
\begin{center}
\includegraphics[width=0.9\columnwidth]{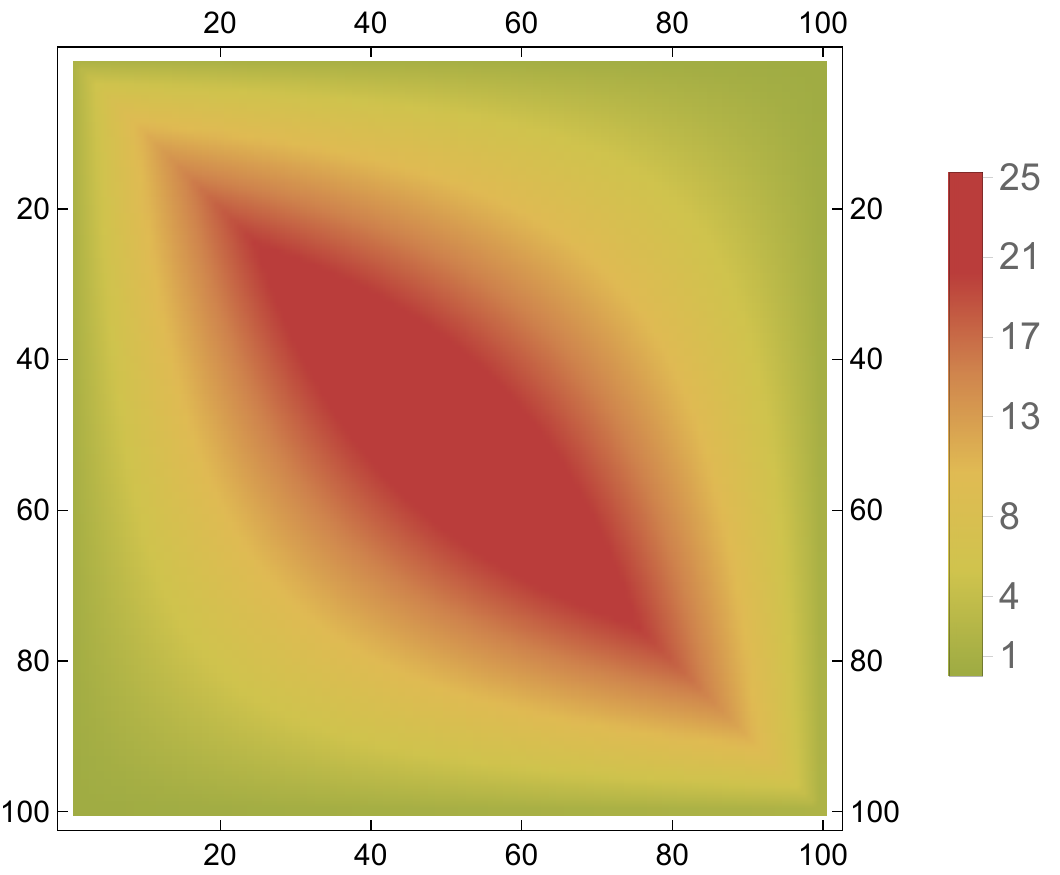}
\end{center}
\caption{A density plot of the covariance matrix for our toy tridiagonal 
precision matrix with $2$ along the diagonals and $-1$ on the first
off-diagonal.}
\label{fig:toycov}
\end{figure}

We consider two examples of covariance (and precision) matrices in this
paper.  First, in order to build intuition, we consider a tridiagonal
precision matrix
$\iC = (\psi_{ij})$ where $\psi_{ii} = 2$, $\psi_{i,i+1} = \psi_{i-1,i} = -1$
and zero otherwise.
The structure of the resulting covariance matrix is shown in
Fig.~\ref{fig:toycov}; unlike the precision matrix,
the covariance matrix has long range support.
The structure of this precision matrix, and the corresponding covariance
matrix, is qualitatively similar to the cosmological examples we will
consider later.  We fix $p=100$; this is similar to the number of
parameters in covariance matrices for current surveys. 

\begin{figure}
\begin{center}
\includegraphics[width=0.9\columnwidth]{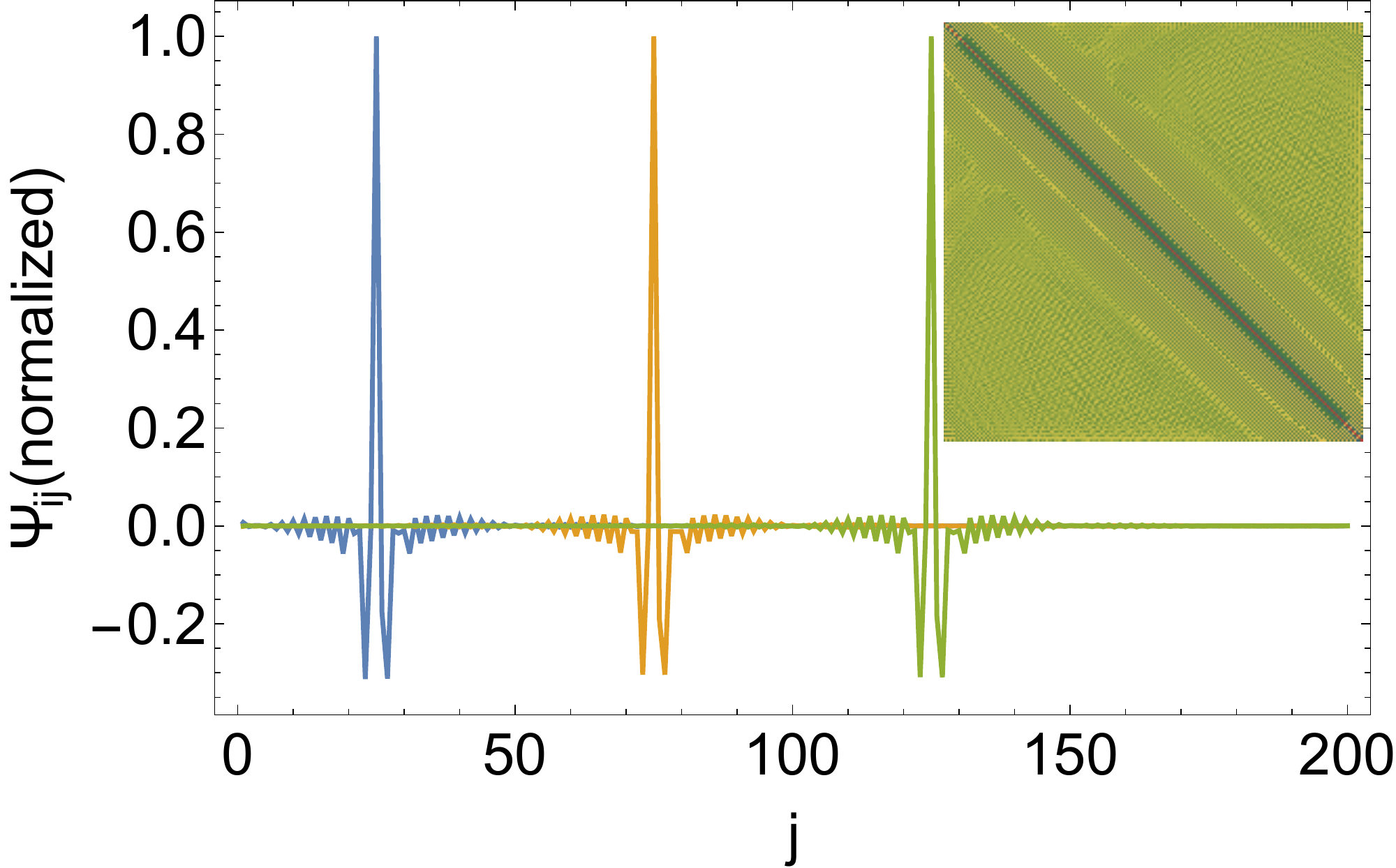}
\end{center}
\caption{A representation of the precision matrix for the cosmologically 
  motivated example. We plot the 25th, 50th, 
  and 75th rows of the matrix, demonstrating that the matrix is clearly
  diagonally dominant, and very sparse. The inset shows that the full matrix
  shows the same structure of the individual rows plotted here.
}
\label{fig:cosmoprecision}
\end{figure}

Second, we use an example more closely related to the covariance matrices
one might expect in galaxy redshift surveys.
Galaxy redshift surveys typically involve negligible measurement
uncertainties, but suffer instead from both sampling variance and
shot noise.
These can be estimated within the framework of a theory, but a
theory for the formation of non-linear objects is not currently
understood.
Instead we use a linear theory approximation.  Specifically, we
compute the covariance matrix of the multipoles of the correlation
function, $\xi_\ell(s)$, assuming Gaussian flucutations evolved
according to linear theory.  The assumed cosmology and power spectrum
are of the $\Lambda$CDM family, with $\Omega_m=0.292$, $h=0.69$,
$n_s=0.965$ and $\sigma_8=0.8$.  The objects are assumed to be linearly
biased with $b=2$ and shot-noise is added appropriate for a number
density of $\bar{n}=4\times 10^{-4}\,h^3{\rm Mpc}^{-3}$.
We evaluate $C_{ij}$ in 100 bins, equally spaced in $s$, for both the
monopole and quadrupole moment of the correlation function, interleaved
to form a $200\times 200$ matrix. Fig.~\ref{fig:cosmoprecision} plots the
corresponding precision matrix. This is clearly dominated by a narrow 
banded structure, and is similar in character to our first case.

\subsection{Entry-wise estimates}

\begin{figure}
\begin{center}
\includegraphics[width=0.9\columnwidth]{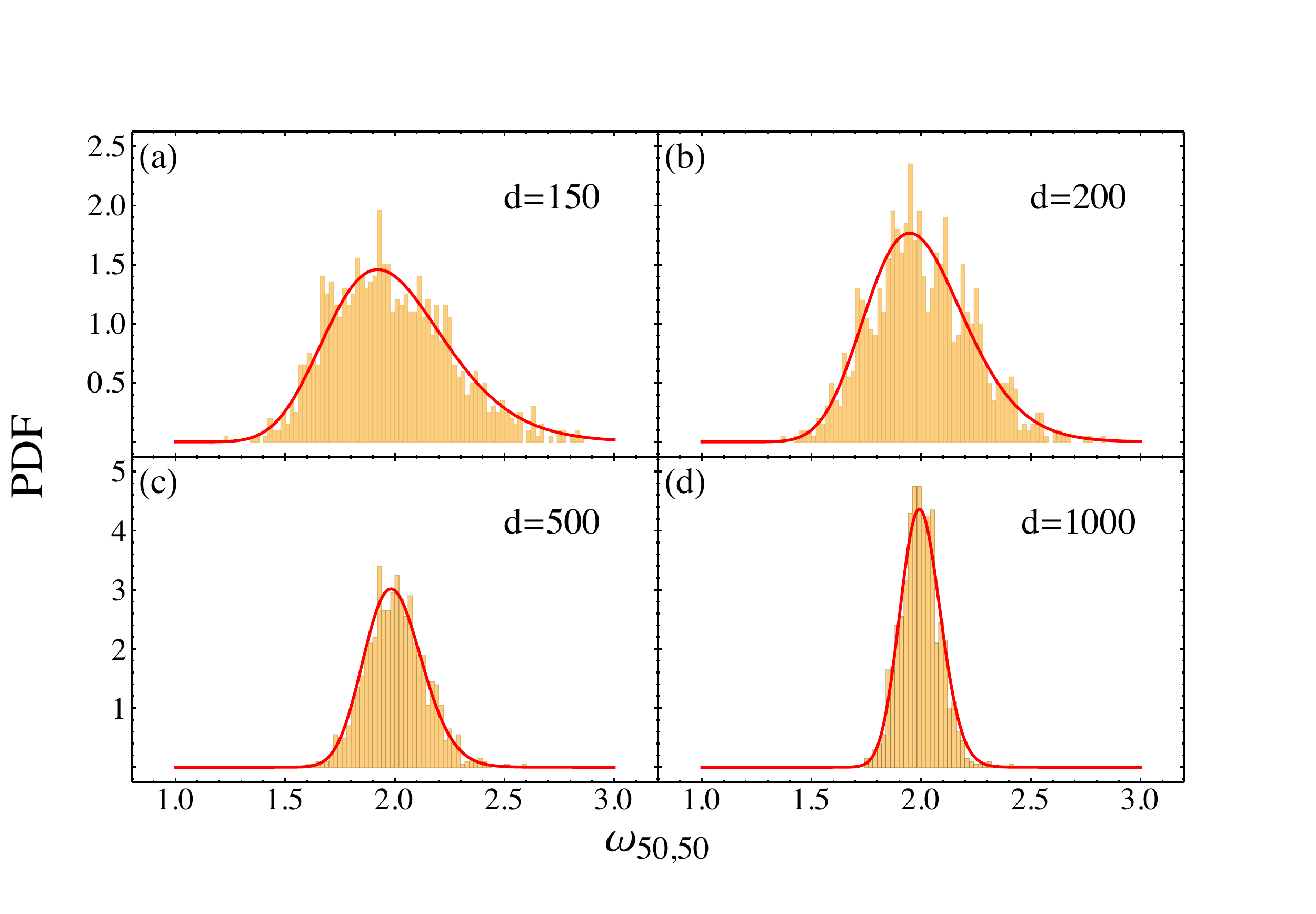}
\end{center}
\caption{Histograms of the recovered values of $\psi_{50,50}$ for different values of $d$, correcting
for the finite sample bias. We assume that the matrix is banded to $k=25$ in all cases.
The solid [red] line is the expected distribution of these values.
}
\label{fig:w5050_k=25_hist}
\end{figure}

\begin{figure}
\begin{center}
\includegraphics[width=0.9\columnwidth]{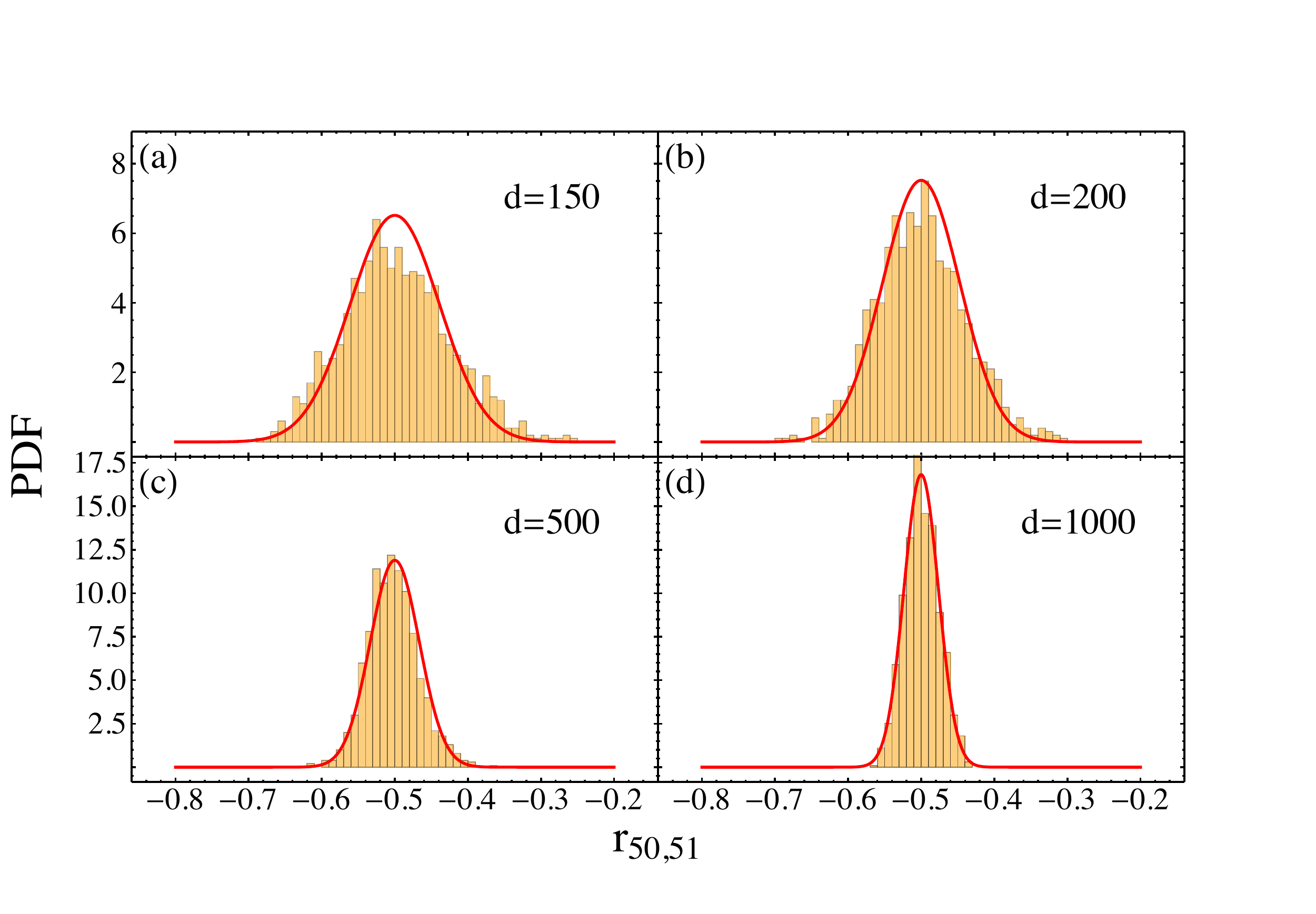}
\end{center}
\caption{Same as Fig.~\ref{fig:w5050_k=25_hist}, except for $r_{50,51}$. 
}
\label{fig:r5051_k=25_hist}
\end{figure}

We start by characterizing the accuracy with which we can estimate individual entries of the
precision matrix. As we discussed previously, we separate out the measurements of the diagonal
elements of $\iC$ from the off-diagonal elements; for the latter, we compute $r_{ij} \equiv 
\psi_{ij}/\sqrt{\psi_{ii} \psi_{jj}}$. Figs.~\ref{fig:w5050_k=25_hist} and \ref{fig:r5051_k=25_hist}
show the distributions of the recovered values for two representative entries of 
the precision matrix of our toy model. The different panels correspond to different numbers
of simulations $d$, while each of the histograms is constructed from an ensemble of 1000 such
realizations. We find good agreement with the theoretically expected distributions of both $\psi_{ii}$
and $r_{ij}$, with the distribution for $r_{ij}$ close to the asymptotically expected Gaussian even
for relatively small numbers of simulations. All of these results assumed a $k=25$ banded structure 
(i.e. 24 non-zero upper/lower off-diagonals); the results for different choices of this banding
parameter are qualitatively similar. Holding the number of simulations $d$ fixed, the scatter in 
the estimates decreases with decreasing $k$, since one is regressing on a smaller number 
of variables in this case.

Given these entrywise estimates (each of which will be individually noisy), we 
turn to the problem of ``smoothing'' away this noise to improve our covariance 
estimates. If one had an a priori model for the structure of the precision matrix, 
one could directly fit to it. For instance, in the case of our toy model, one 
might use the fact that the matrix has constant off-diagonals. However, in general,
one might only expect the matrix to be smooth (nearby entries with the same 
value); in such a case, using eg. smoothing splines could provide a reasonably 
generic solution.

\begin{figure}
  \begin{center}
    \includegraphics[width=0.9\columnwidth]{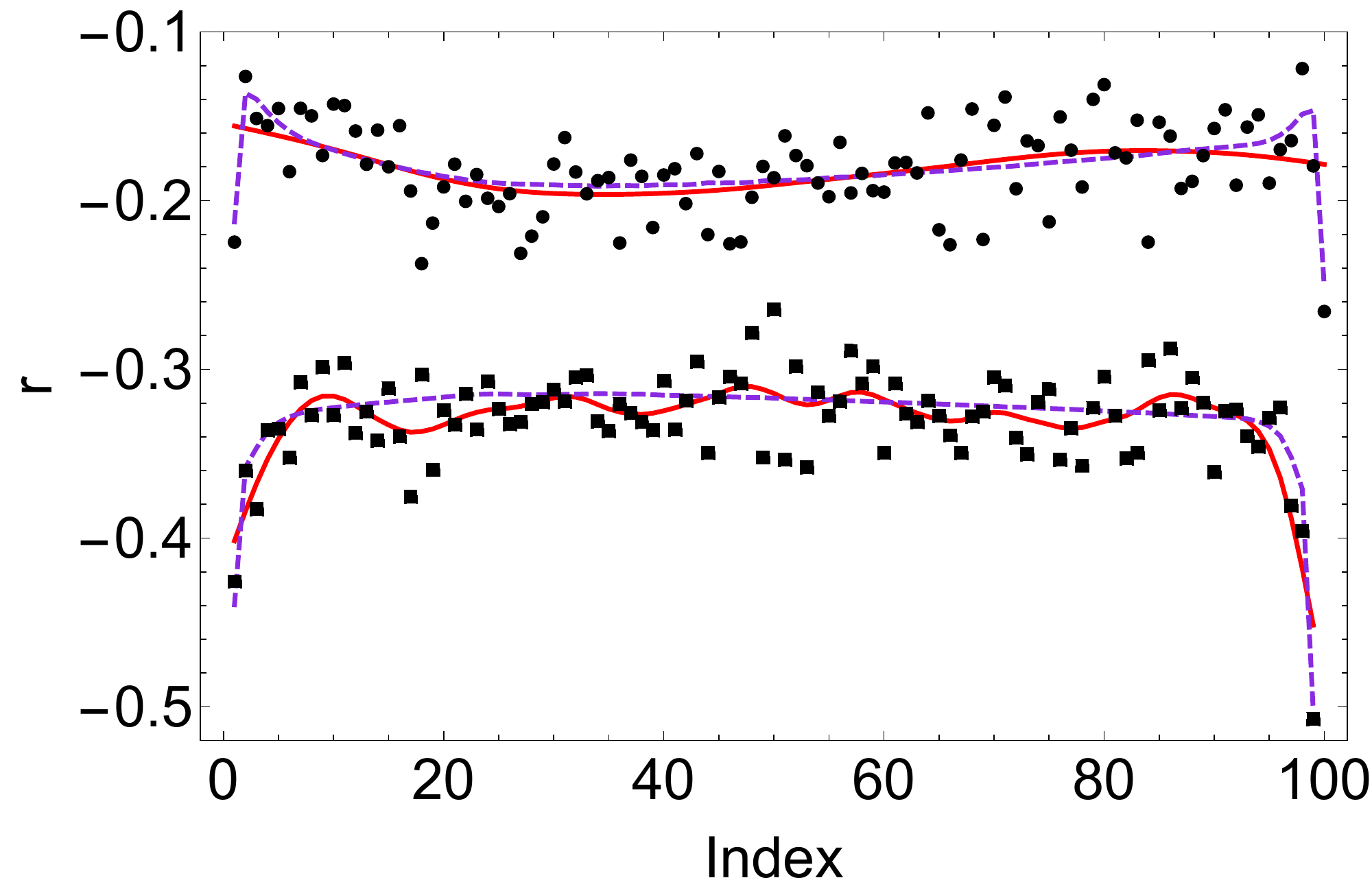}
  \end{center}
  \caption{A demonstration of the impact of smoothing on the elements
    of the precision matrix for the cosmological model. We plot the odd
    elements of the first off-diagonal [top, circles] and the even elements of
    the second off-diagonal for our cosmological model [bottom, squares] (see
    text for details for how the covariance matrix is packed).  The precision
    matrix was estimated by assuming $k=20$ and $d=2000$ simulations.  The
    points are the raw entry-wise estimates, the dashed line is the true value,
    while the solid line is the smoothed version. Note that smoothing can
    significantly reduce the variance of the estimates. More complicated
  smoothing schemes could reduce the bias at the very edges.  }
  \label{fig:smoothDiagonal}
\end{figure}

As a worked non-trivial example, we consider smoothing an estimate of our
cosmological model. Given that the underlying galaxy correlation function 
is a smooth function, we would expect the off-diagonals of the precision 
matrix to be smooth. However, given that our matrix was constructed by 
interleaving the monopole and quadrupole correlation functions, we'd expect 
this interleaving to persist in the off-diagonals. We therefore smooth
the odd and even elements of each off-diagonal separately. Fig.~\ref{fig:smoothDiagonal}
shows the results for two example minor diagonals, comparing it to the 
raw entrywise estimates as well as the true value. For the majority of the points, 
smoothing significantly reduces the noise in the entrywise estimates.  
It can also miss features in the model, if those variations are smaller than the
noise. We observe this effect at the edges of the top curves, where the smoothed
curves miss the drop-off for the initial and final points. This is clearly a function
of the relative sizes of these effects, as is evident from the lower sets of curves
where the smoothed estimates track the variations at the edges better.
  
\begin{figure}
  \begin{center}
    \includegraphics[width=0.9\columnwidth]{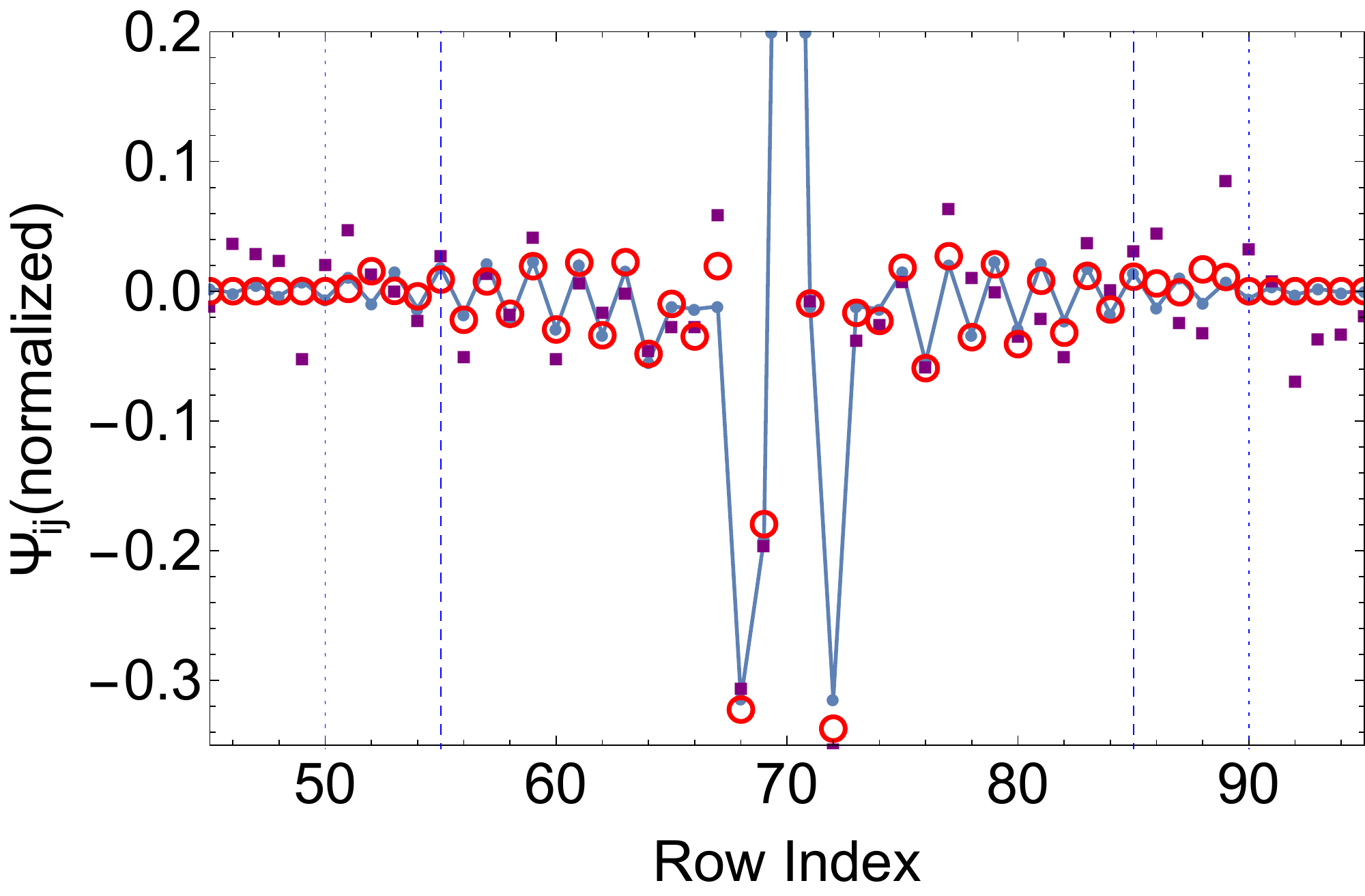}
  \end{center}
  \caption{A part of the 70th row of the cosmological precision matrix,
    normalized by $\sqrt{\psi_{ii} \psi_{jj}}$.  The filled circles connected by
    solid lines is the true precision matrix, the filled squares show the sample
    precision matrix, while the open circles are the estimate developed in this
    paper. The precision matrix was estimated from $d=1000$ simulations, and we
  assume a banding of $k=20$ to estimate the matrix (short dashed lines). The fiducial 
  $k=15$ banding that we use for this matrix 
  is marked by the long dashed lines.  }
  \label{fig:cosmo_rowplot}
\end{figure}

Our final step is to ensure a positive definite precision matrix using 
the algorithm presented in the previous section. Fig.~\ref{fig:cosmo_rowplot} 
shows the final results, plotting an example row from the cosmological case. 
Our estimator tracks the oscillating structure in the precision matrix, and 
is clearly less noisy than the direct sample precision matrix. 
This improvement directly translates into improved performance of the precision
matrix.

\subsection{Selecting the banding}

\begin{figure}
  \begin{center}
    \includegraphics[width=0.9\columnwidth]{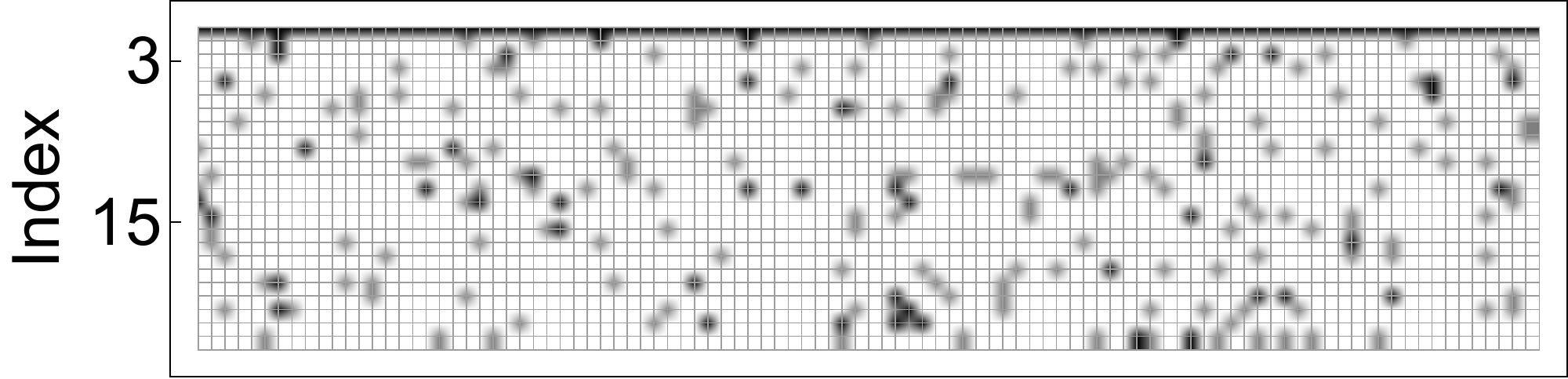}
    \includegraphics[width=0.9\columnwidth]{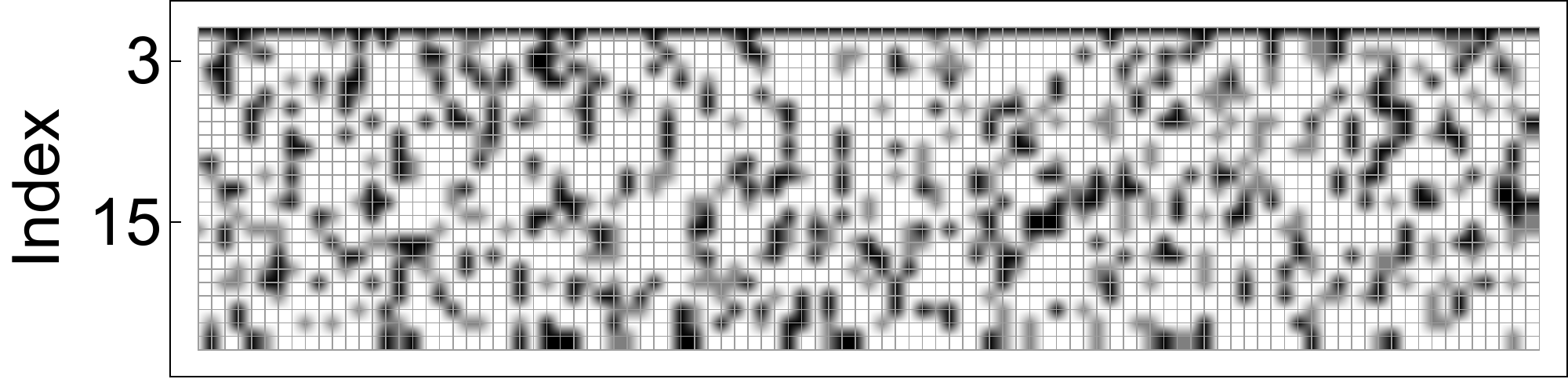}
  \end{center}
  \caption{Selecting the banding for the toy model, with each row 
    representing whether a particular off-diagonal was consistent with
    zero or not. 
    Each row represents an off-diagonal, with the index of the off-diagonal on 
    the $y$-axis, while each column represents a set of $d=500$ simulations used to estimate
    the precision matrix. In order to estimate the precision matrix, we assume a banding of
    $k=25$.
    Unfilled boxes show diagonals consistent with zero, 
    gray boxes show diagonals that are inconsistent with zero at the 95\% level but not at 99\%, 
    while black represents diagonals inconsistent with zero at the 99\% level. An 
    index of one represents the first off-diagonal.
    The top panel considers the unsmoothed 
    matrix, while the lower panel considers the smoothed case. The tridiagonal nature of the matrix 
    is clearly apparent here, with the first off-diagonal clearly nonzero, and no structure evident for the remaining
    cases. We choose $k=3$ as our fiducial choice for both cases.
  }
  \label{fig:toy_ktest}
\end{figure}

We now turn to the problem of determining the banding parameter $k$. Our goal here isn't
to be prescriptive, but to develop a simple guide. We anticipate that 
the choosing an appropriate banding will depend on the particular problem as well as
some numerical experimentation.

\begin{table}
  \centering
  %% See ktesttable.nb for the code...
%% This is just hardcoded, sorry!
\begin{tabular}{ccc}
\hline
$m$ & Unsmoothed & Smoothed \\
\hline
70 & 2.47 & 0.75 \\
80 & 2.52 & 0.71 \\
90 & 2.56 & 0.67 \\
100 & 2.59 & 0.63 \\
\hline
\end{tabular}

  \caption{An example of the thresholds at the 95\% level for the unsmoothed and
  smoothed off-diagonals (see text for details on how these are computed). While the thresholds
  increase for the unsmoothed case, they decrease for the smoothed case since the increase in the 
  number of points better constrain the spline.}
  \label{tab:thresh}
\end{table}

Since we do not expect the off-diagonals of the precision matrix to be exactly
zero, choosing a banding reflects a trade-off between bias and variance.
By setting small off-diagonals to be zero, our estimated precision matrix
will, by construction, be biased.
However, any estimate of these off-diagonal elements from a finite number
of simulations will be dominated by the noise in the measurements.
This also implies that the banding will, in general depend on the number
of simulations (except in the case that the matrix is truly sparse);
we will see this explicitly in the next section.

We propose a simple thresholding scheme to determine whether an entire
off-diagonal is consistent with zero or not - we set an off-diagonal to
zero if the maximum absolute value of all its elements is less than a
pre-determined threshold.
For the unsmoothed estimate of the precision matrix, we determine this
threshold as follows. 
Assume an off-diagonal has $m$ elements $x_i$, all drawn from a Gaussians
with mean zero and a known variance (but with arbitrary correlations). 
Then, the probability of exceeding a threshold value $X$ can be 
bounded by the following 
\begin{align}
  P({\rm max}\ x_i > X) \le \sum_i P(x_i > X) = m P(x > X)
\end{align}
where the last $P(x > X)$ is the complementary cumulative Gaussian probability distribution. 
Choosing an appropriate failure\footnote{mis-classifying a zero off-diagonal as non-zero}
rate $m P(x > X)$ determines the appropriate threshold to use. 

We follow a similar procedure for the smoothed estimate of the precision matrix, although
the choice of the threshold is complicated by the smoothing procedure and the correlations
between different elements on the same off-diagonal. We therefore estimate the threshold 
by Monte Carlo, simulating $m$ random Gaussian variables, fitting them with a cubic spline
and then tabulating the distribution of maximum values. This process ignores the correlations
between the points and so, we expect our estimates to only be approximate. Table~\ref{tab:thresh}
shows an example of these thresholds as a function of $m$.

\begin{figure}
  \begin{center}
    \includegraphics[width=0.9\columnwidth]{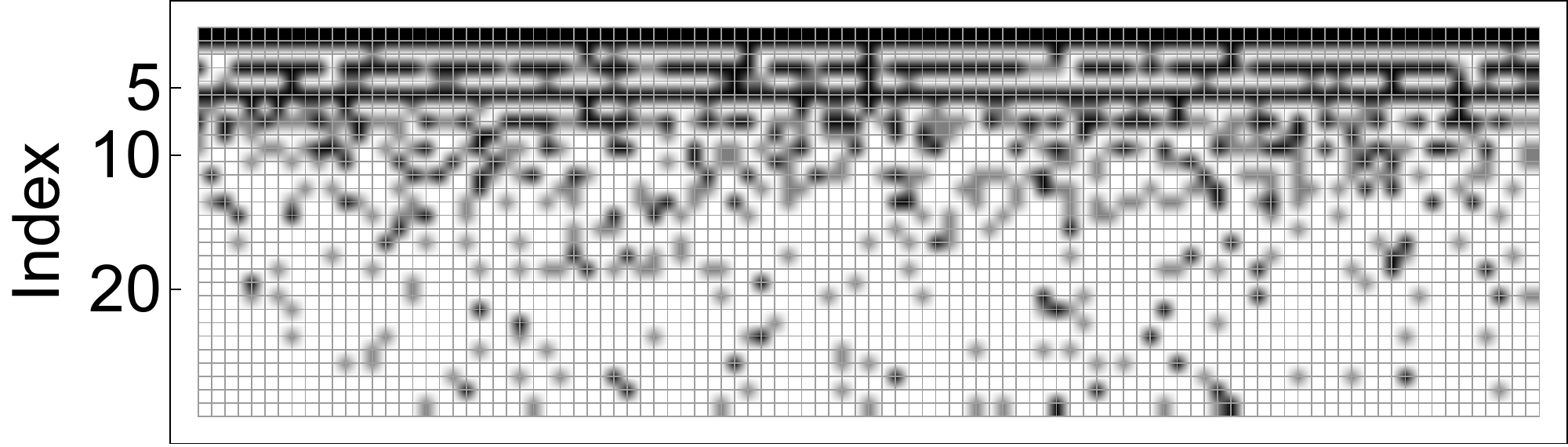}
    \includegraphics[width=0.9\columnwidth]{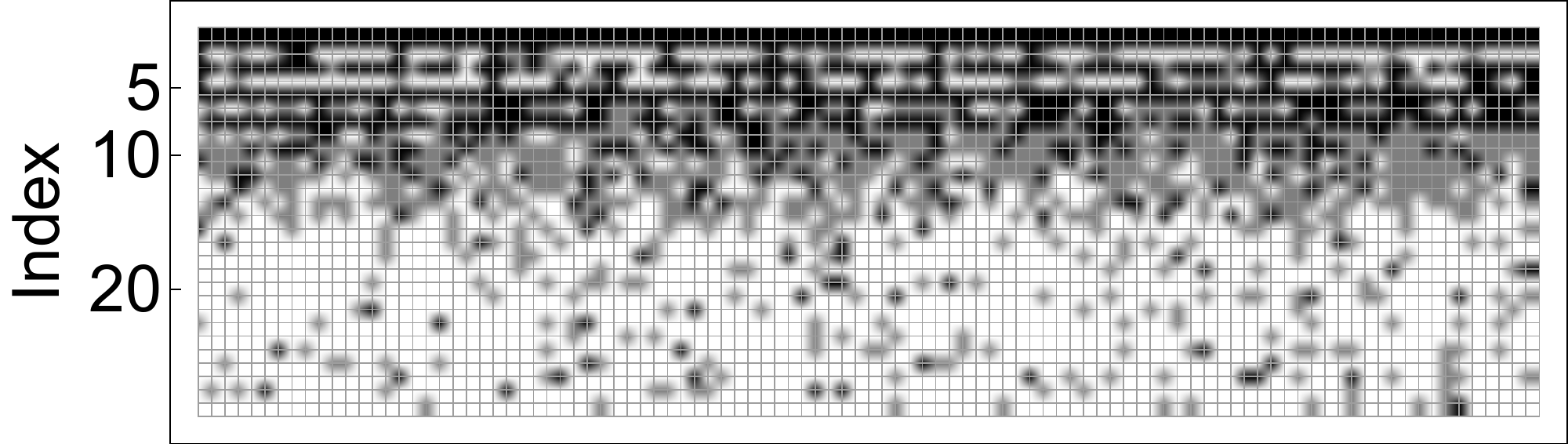}
  \end{center}
  \caption{Same as Fig.~\ref{fig:toy_ktest} except for the cosmological model. 
    The number of realizations per column is $d=1000$, and the banding assumed for
    in the estimate is $k=30$.
    The difference 
    between the smoothed and unsmoothed cases is more apparent here.
    Smoothing requires us to estimate more off-diagonals, due to the reduction
    in the noise.
    We choose $k=10$ and $k=15$ for 
  the unsmoothed and smoothed cases respectively.}
  \label{fig:cosmo_ktest}
\end{figure}

Figs.~\ref{fig:toy_ktest} and ~\ref{fig:cosmo_ktest} plot, for a set of simulations, which 
off-diagonals are determined to be non-zero, using the above criteria. 
Each row of these figures corresponds to an off-diagonal, whereas each column represents
an independent set of simulations from which a precision matrix can be estimated. Since our 
procedure assumes a banding of the precision matrix, we start by selecting a conservative 
choice of the band. Filled boxes show off-diagonals that are inconsistent with zero, with 
the shading representing the confidence level for this. A banded matrix in this figure
would appear as a filled set of rows.

We start with the unsmoothed cases first (upper panels). The tridiagonal nature of the
toy precision matrix is clear, with only a single row clearly visible. The cosmological 
example is less sharp, but a central band is still clear. The smoothed cases (lower
panels) are noisier, because our estimates of the thresholds ignored correlations.
Even with this increased noise, we discern no trend suggesting an increased banding
for the toy example. For the cosmological example however, the non-zero band is clearly
increased, implying that one must work out to a larger band. This is to be expected, 
since the elements of the precision matrix is this case are not exactly zero. Reducing
the noise (by smoothing) reduces the threshold below which one may estimate an element 
of the precision matrix to be zero. We find a similar trend with increasing numbers
of simulations (discussed further below).

A shortcoming of the above is the lack of an objective, automatic determination of the banding.
We do not have such a prescription at this time (and it isn't clear that such a prescription 
is possible in general). We therefore advocate some level of numerical experimentation when 
determining the appropriate band.

\subsection{Performance of estimates}

We now turn to the performance of our precision matrix estimates as a function of the
number of simulations input. There are a number of metrics possible to quantify
``performance''. The most pragmatic of these would be to propagate the estimates
of the precision matrix through parameter fitting, and see how the errors in the 
precision matrix affect the errors in the parameters of interest. The disadvantage of 
this approach is that it is problem-dependent and therefore hard to generalize. 
We defer such studies to future work that is focused on particular applications. 

A more general approach would be to quantify the ``closeness'' of our precision matrix
estimates to truth, using a ``loss'' function.
There are a variety of ways to do this, each of which test different
aspects of the matrix. We consider five such loss functions here :
\begin{enumerate}
  \item {\it Frobenius Norm} :
    \begin{align}
      || \Delta \iC ||_F &\equiv || \iC - \iCh ||_F
    \end{align}
    This is an entrywise test of the elements of the precision matrix, and 
    is our default loss function. Clearly, reducing the Frobenius norm
    will ultimately improve any estimates derived from the precision matrix, 
    but it is likely not the most efficient way to do so. In the basis where
    $\Delta \iC$ is diagonal, the Frobenius norm is just the RMS of the 
    eigenvalues, and can be used to set a (weak) bound on the error on $\chi^2$.
  \item {\it Spectral Norm} :
    \begin{align}
      || \Delta \iC ||_2 &\equiv || \iC - \iCh ||_2
    \end{align}
    This measures the largest singular value (maximum absolute eigenvalue) of $\Delta \iC$.
    This yields a very simple bound on the error in $\chi^2$ --- $||\Delta \iC||_2 |\x|^2$ where
    $x$ is the difference between the model and the observations.
  \item {\it Inverse test} : 
    A different test would be to see how well $\iCh$ approximates the true inverse of $\C$.
    A simple measure of this would be to compute $|| \C \iCh - I ||_F$ = 
    $|| \C \iCh - \C \iC||$ = $|| \C \Delta \iC||$. However, this 
    measure is poorly behaved. In particular it isn't invariant under transposes, 
    although one would expect $\iCh$ to be an equally good left and right inverse. To
    solve this, we use the symmetrized version $||\C^{1/2} \iCh \C^{1/2} - I||_F$, although
    for brevity, we continue to denote it by $||\C \Delta \iC||_F$.
  \item {\it $\chi^2$ variance} :
    Given that parameter fits are often done by minimizing a $\chi^2$ function, we 
    can aim to minimize the error in this function due to an error in the precision
    matrix. If we define $\Delta \chi^2 = \x^{t} \Big( \Delta \iC \Big) \x$, where 
    as before, $\x$ is the difference between the data and the model, we define 
    the $\chi^2$ loss as the RMS of $\Delta \chi^2$. In order to compute this, we
    need to specify how $\x$ is distributed. There are two options here. The 
    first comes from varying the input parameters to the model, while the second 
    comes from the the noise in the data. The former is 
    application-dependent and we defer specific applications to future work. The 
    second suggests $\x \sim \normal{0}{\C}$, in which case we find
    \begin{align}
      \sigma\Big(\Delta \chi^2\Big) &= \Bigg[ 2 {\rm Tr} \big(\Delta \iC \C \Delta \iC \C\big)
    + \big( {\rm Tr} \iC \C \big)^2 \Bigg]^{1/2}
    \end{align}
  \item {\it Kullback-Leibler (KL) divergence} :
    Our final loss function is the Kullback-Leibler divergence\footnote{A
    divergence is a generalization of a metric which need not be symmetric
    or satisfy the triangle inequality.
    Formally a divergence on a space $X$ is a non-negative function on the
    Cartesian product space, $X\times X$, which is zero only on the diagonal.}
    between $\C$ and $\iC$, defined by 
    \begin{align}
      KL &\equiv \frac{1}{2} \Bigg[ {\rm Tr} (\C \iC) - {\rm dim}(C) - \log {\,\rm det} (C \iC) \Bigg]\,.
    \end{align}
    The Kullback-Leibler divergence can be 
    interpreted as the expectation of the Gaussian likelihood ratio of $\x$ computed with the
    estimated and true precision matrices. As in the inverse variance case, we 
    assume $\x \sim \normal{0}{\C}$, which captures the variation in the data.
\end{enumerate}

\begin{figure}
\begin{center}
  \includegraphics[width=0.9\columnwidth]{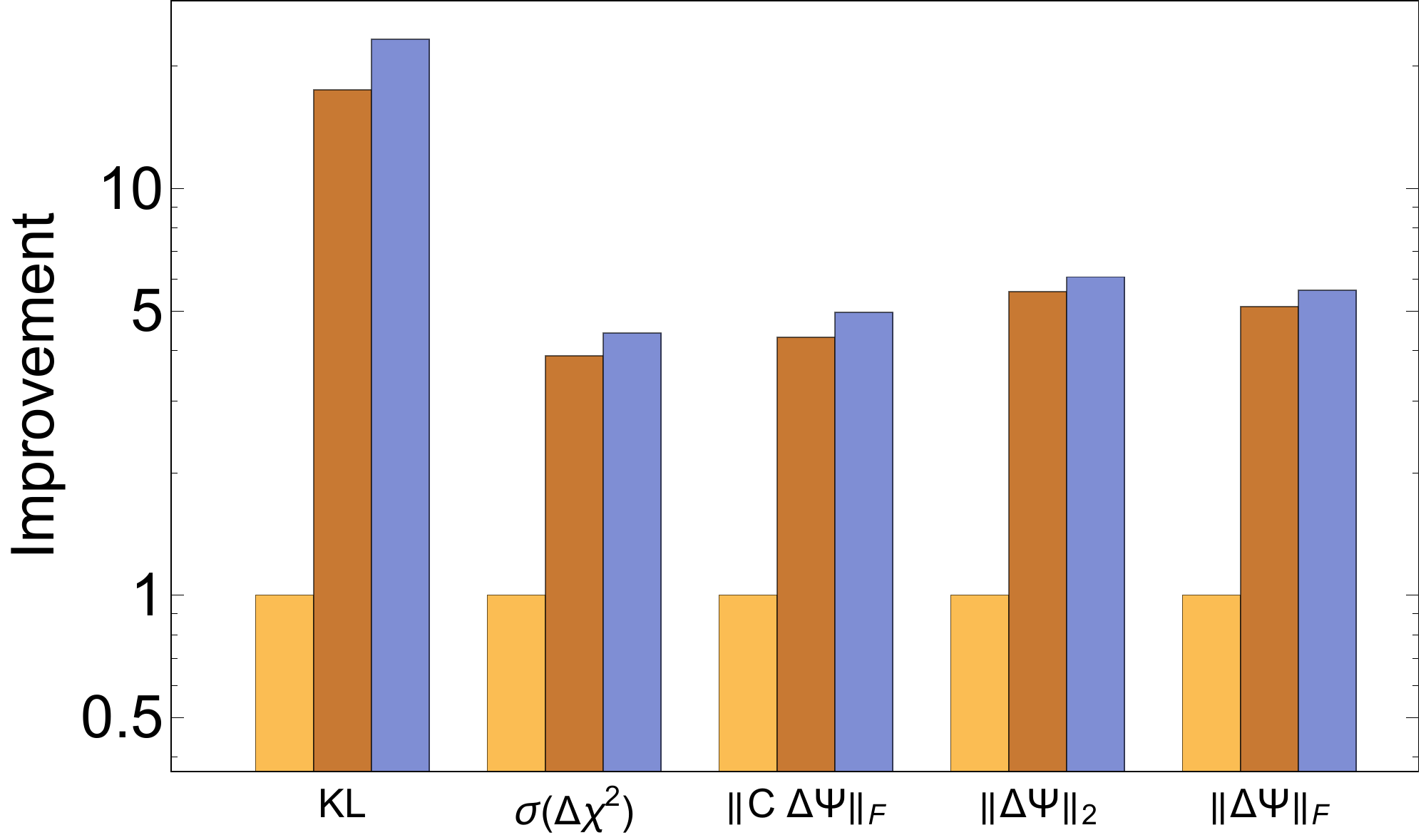}
\end{center}
\caption{The improvement over the sample precision matrix ${\rm Loss(sample)}/{\rm Loss}$
for different norms for
the toy tridiagonal precision matrix.
From left to right, each group  plots the improvement for 
the sample precision matrix (1 by construction), the unsmoothed and smoothed
precision matrices. We assume a $k=3$ banding in both the unsmoothed and smoothed
cases, and all cases assume $d=500$. We average over 50 such realizations in this 
figure, although the improvements for individual realizations are similar.
}
\label{fig:toynormbars}
\end{figure}

\begin{figure}
\begin{center}
  \includegraphics[width=0.9\columnwidth]{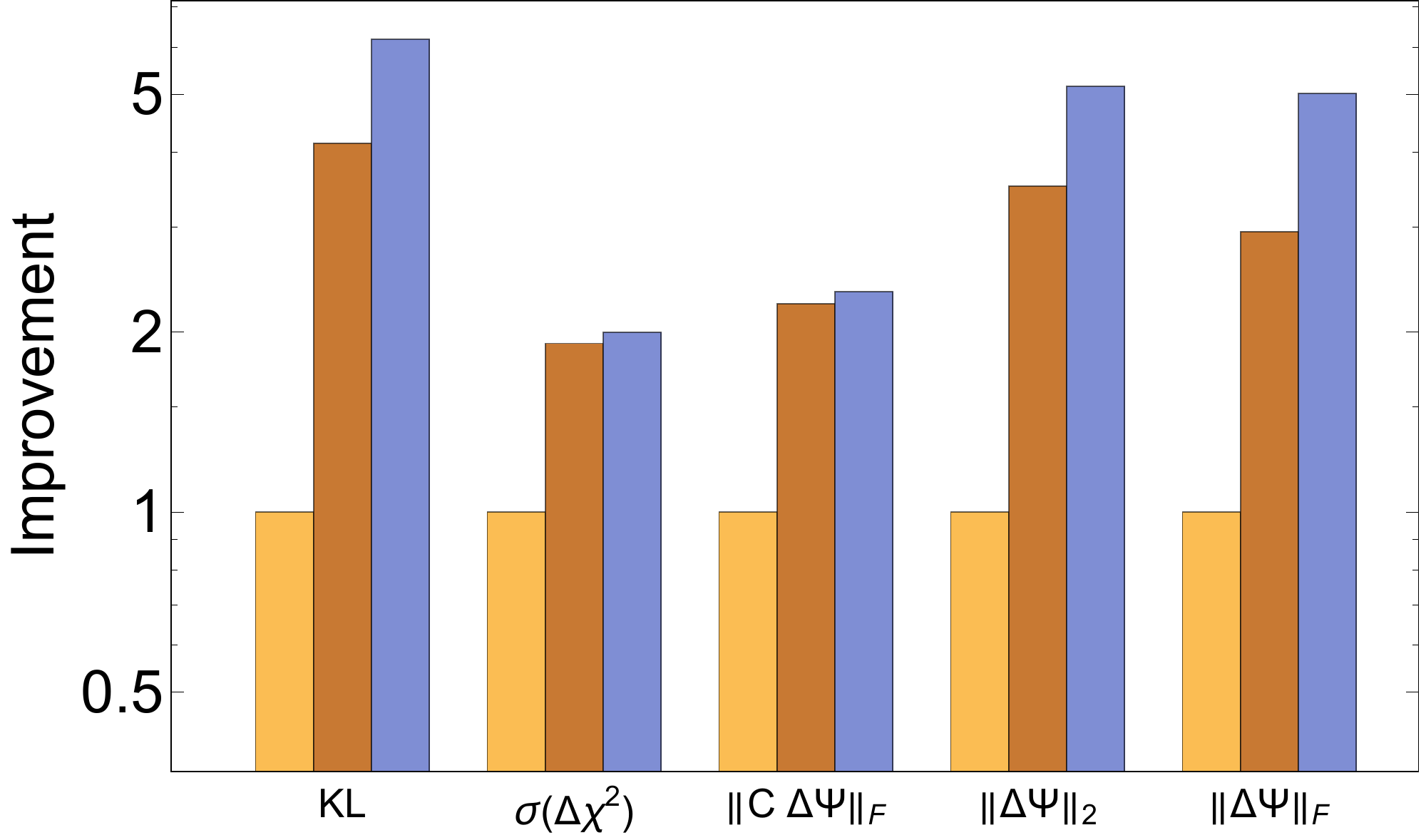}
\end{center}
\caption{The same as Fig.~\ref{fig:toynormbars} but for the cosmological model.
  The unsmoothed covariance matrix assumes $k=10$, while the smoothed covariance
  matrix uses $k=15$. All three cases use $d=1000$.
}
\label{fig:cosmonormbars}
\end{figure}

Figures~\ref{fig:toynormbars} and \ref{fig:cosmonormbars} show these different losses
for our toy and cosmological models; we plot the ratio of the loss obtained using 
the sample precision matrix to the loss obtained using the techniques presented in this
paper.
The figures show the improvement averaged over fifty
realizations, although the improvement for individual realizations is similar. For all
choices of a loss function, we find that the techniques presented here yield a 
factor of a $\sim$few improvement over simply inverting the sample precision matrix.
The largest gains come from exploiting the sparsity of the precision matrix and
directly estimating it from the simulations. The secondary smoothing step yields 
a smaller relative (although clear) improvement. Given the similar behaviour
of all the loss functions, we focus on the Frobenius norm below for brevity.

\begin{figure}
  \begin{center}
    \includegraphics[width=0.9\columnwidth]{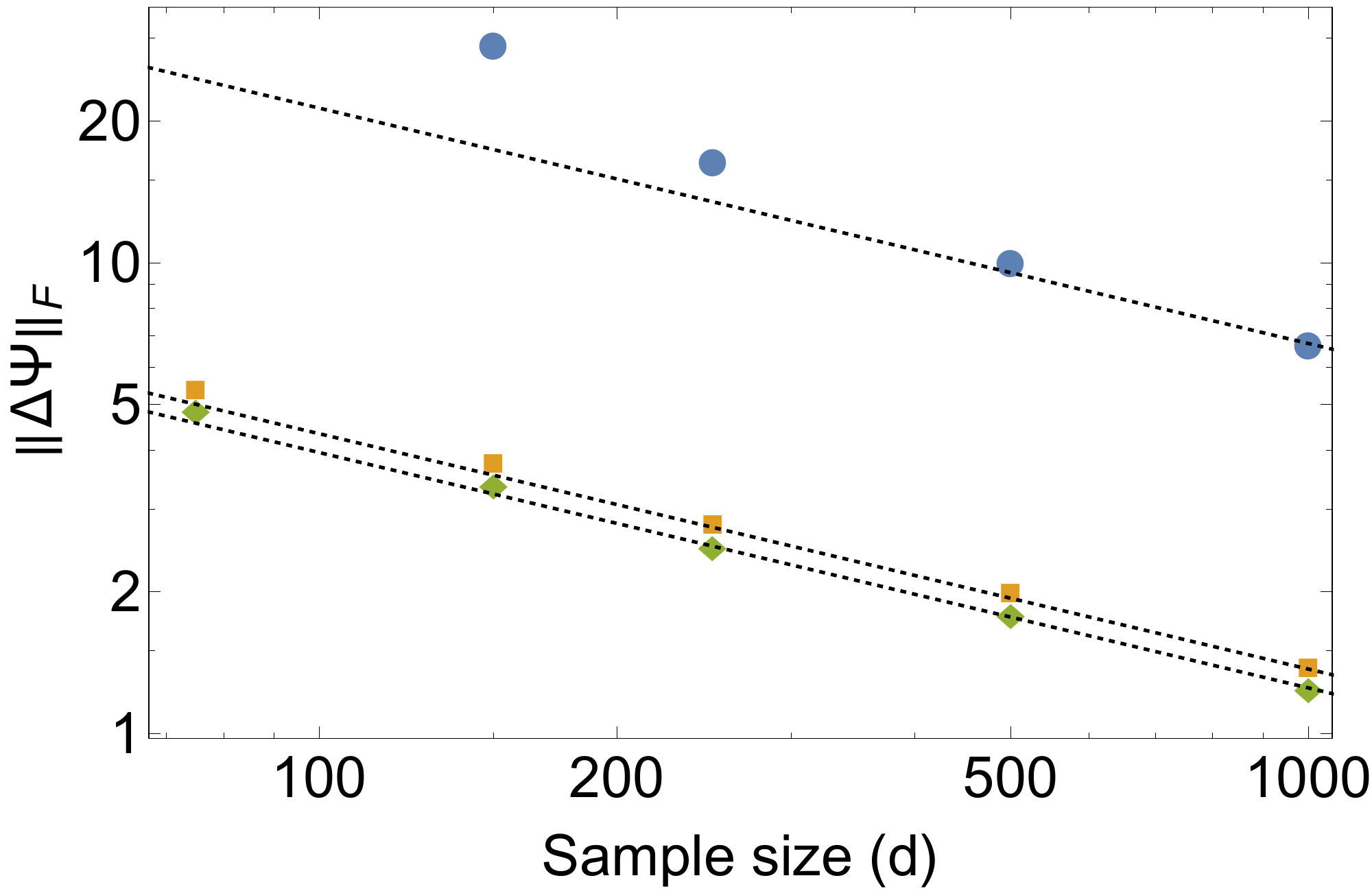}
  \end{center}
  \caption{The Frobenius loss $|| \Delta \iC||_F$ for our toy model as a function of sample
    size $d$ for the sample precision matrix (blue circles), our estimate of the
    precision matrix with and without the intermediate smoothing step (green
    diamonds and orange squares respectively). The latter two cases assume
    a banding of $k=3$. The dashed lines show a $d^{-1/2}$ trend which is more
    quickly attained by the estimates presented in this work than by the 
    sample precision matrix. Recall that this is a $100\times 100$ matrix -
    one cannot estimate the sample precision matrix with $d<100$. This
    restriction isn't present for the estimator presented here.
  }
  \label{fig:toynormd}
\end{figure}

\begin{figure}
  \begin{center}
    \includegraphics[width=0.9\columnwidth]{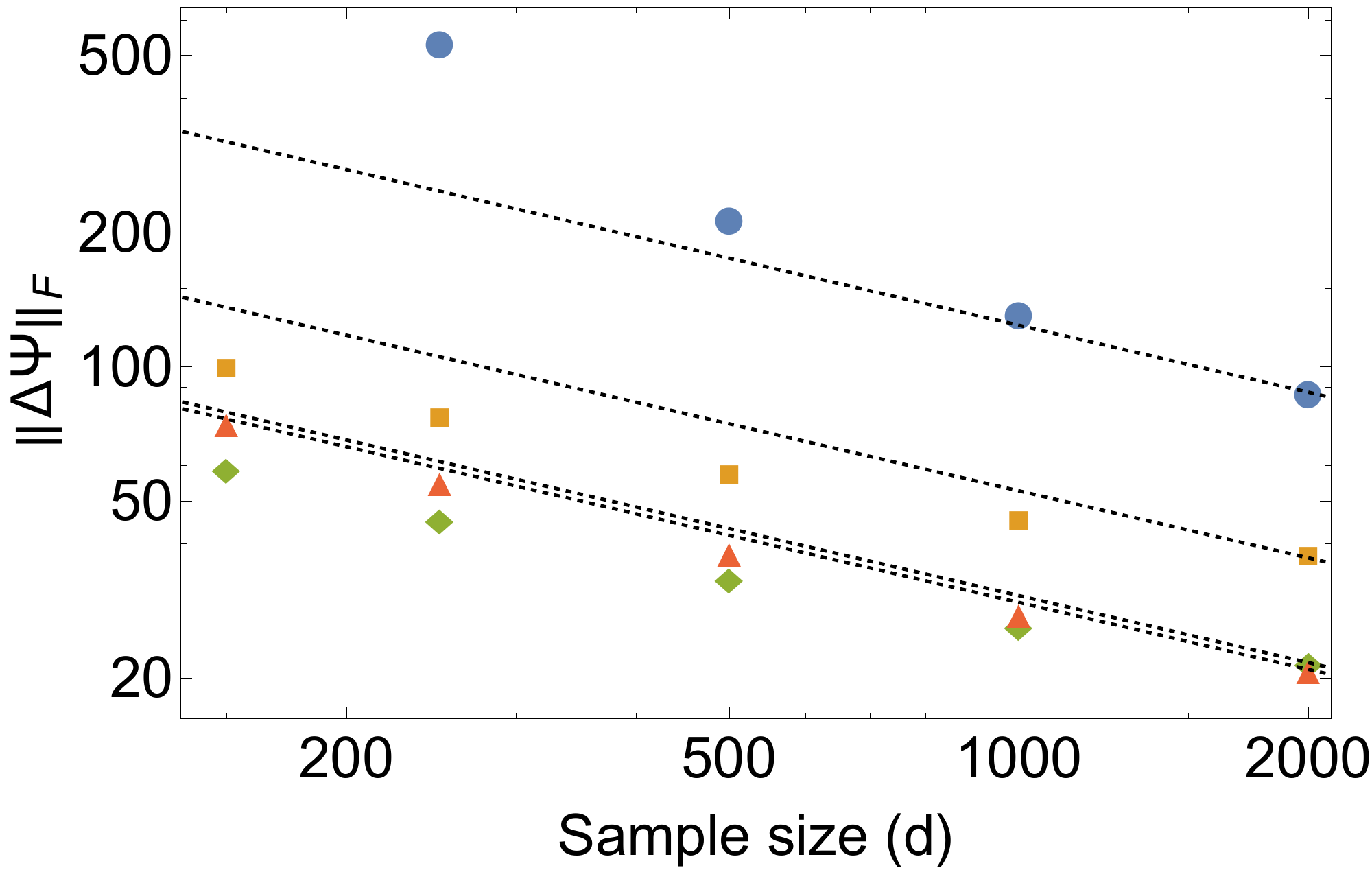}
  \end{center}
  \caption{Analogous to Fig.~\ref{fig:toynormd} except now for the 
    cosmological model. As in the previous case, the circles [blue] show 
    the sample precision matrix, squares [orange] - our unsmoothed estimate 
    with $k=10$, diamonds [green] - our smoothed estimate with $k=15$, and
    and triangles [red] - our smoothed estimate with $k=30$. Unlike the 
    toy model, we see our estimators flatten out with increasing $d$ at fixed
    values of $k$. This reflects the fact that the precision matrix here is not
    strictly sparse, and that with increasing values of $d$, one can reliably
    estimate more elements of the matrix.
  }
  \label{fig:cosmonormd}
\end{figure}

We now turn to how the loss depends on the number of simulations $d$; 
Figs.~\ref{fig:toynormd} and ~\ref{fig:cosmonormd} summarize these 
results for both models considered here. The improvements seen above 
are immediately clear here. At a fixed number of simulations, one 
obtains a more precise estimate of the precision matrix; conversely, 
a significantly smaller number of simulations is required to 
reach a target precision. We also note that we can obtain estimates 
of the precision matrix with $d < p$ simulations. Given that we assume 
the precision matrix is sparse, this isn't surprising, although we
re-emphasize that the usual procedure of first computing a 
covariance matrix and then inverting it doesn't easily allow 
exploiting this property. 
%%Interestingly, for both examples we consider
%%here, this small number of simulations yields a more accurate estimate
%%of the precision matrix that what we obtain for the largest 
%%number of simulations using the sample precision matrix. 

Analogous to the fact that the sample 
precision matrix requires $d>p$ to obtain an estimate, our
approach requires $d > k$ to perform the linear regressions. 
The gains of the method come from the fact that $k$ can be 
significantly smaller than $d$.

We can also see how the accuracy in the precision matrix scales 
with the number of simulations. For $d \gg p$, the error scales 
as $d^{-1/2}$ as one would expect from simple averaging. However, 
as $d$ gets to within a factor of a few of $p$, the error deviates
significantly from this scaling; at $d<p$ the error is formally 
infinite, since the sample covariance matrix is no longer invertible.

The $d$ dependence for our estimator is more involved. For the 
case of the toy model, we find that the error lies on the 
$d^{-1/2}$ scaling for the range of $d$ we consider here, with 
the smoothed estimator performing $\sim$10\% better. Note that we 
do not converge faster than $d^{-1/2}$ - that scaling is set 
by the fact that we are still averaging over simulations - 
but the prefactor is significantly smaller. At fixed banding, 
the cosmological model starts close to a $d^{-1/2}$ scaling, but then
flattens off as $d$ increases. This follows from the fact that the
true precision matrix is not strictly zero for the far off-diagonals,
and therefore, the error in the estimator has a minimum bound. 
However, the appropriate banding will be a function of $d$, since
as the number of simulations increases, one will be able to estimate
more off-diagonal terms. We see this explicitly in the figure 
where the $k=30$ curve starts to outperform the $k=15$ curve at large $d$.

\begin{figure}
  \begin{center}
    \includegraphics[width=0.9\columnwidth]{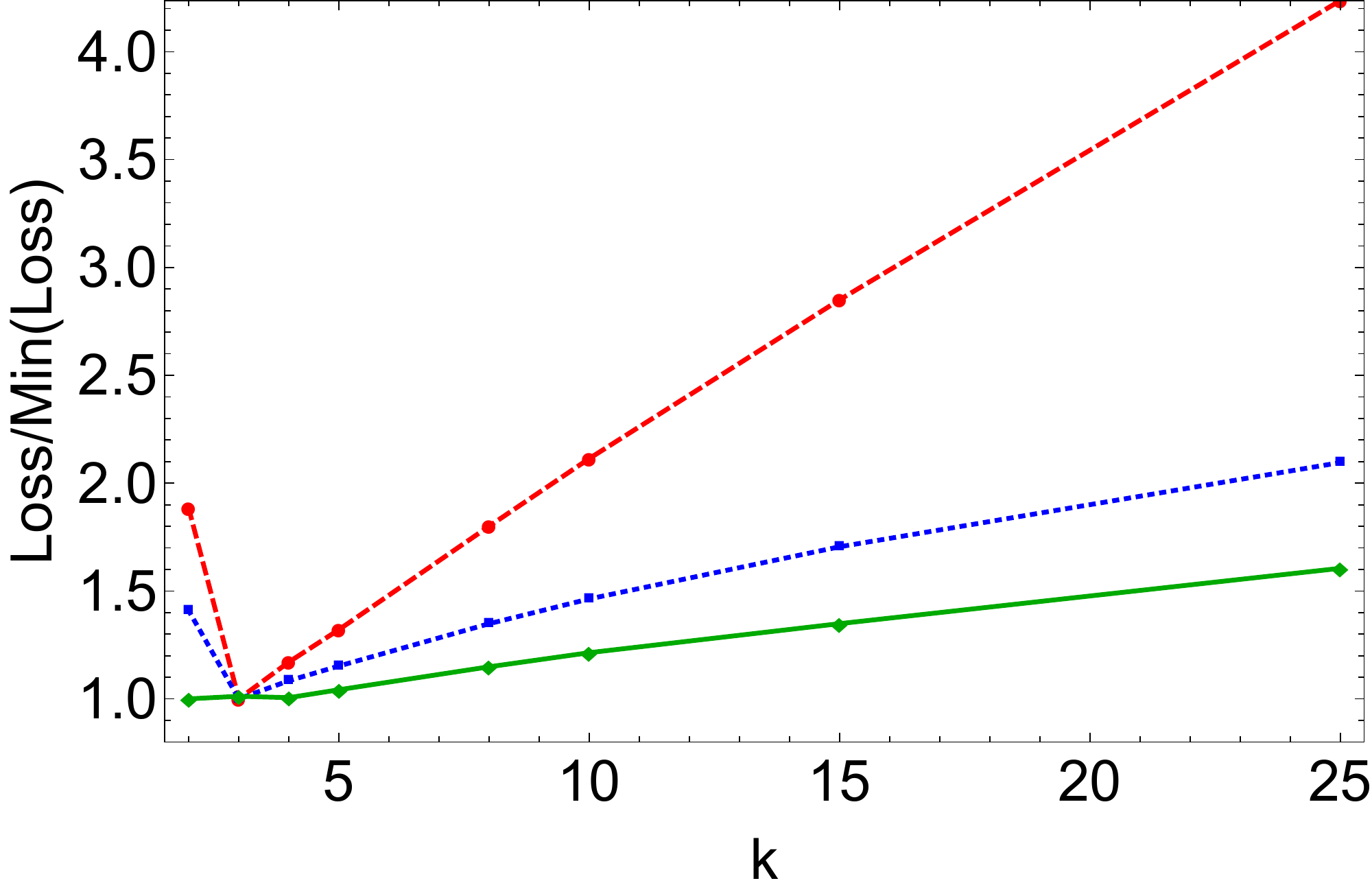}
  \end{center}
  \caption{Matrix losses as a function of the choice of the banding parameter
    $k$ for the toy model. The dashed [red], dotted [blue] and solid [green]
    lines correspond to the KL, $||C \Delta \Omega||$ and $|| \Delta
    \Omega||_{F}$ losses respectively.  The losses are all normalized to the
    minimum value of the loss. Recall that $k$ is defined including the main
    diagonal (eg. $k=2$ corresponds to one-off diagonal element). All of these
    are computed for $d=500$ simulations. We find a well-defined minimum here,
    although for two of the three norms, it is at $k=3$ and not $k=2$.
  }
  \label{fig:loss_toy}
\end{figure}

Motivated by this, we consider how the various losses vary as a function
of the assumed banding at fixed number of simulations. For the toy model, 
we find a well defined minimum $k$ value. Intriguingly, except for the 
Frobenius norm, it is one larger that the true value. For the cosmological 
case, the answer is less clear. All the norms suggest $k>10$, but have 
very weak $k$ dependence after that, with possible multiple maxima. 
This suggests that the particular choice of $k$ will tend to be 
application specific (and require some numerical experimentation). 
We defer a more careful study of this to later work.

\begin{figure}
  \begin{center}
    \includegraphics[width=0.9\columnwidth]{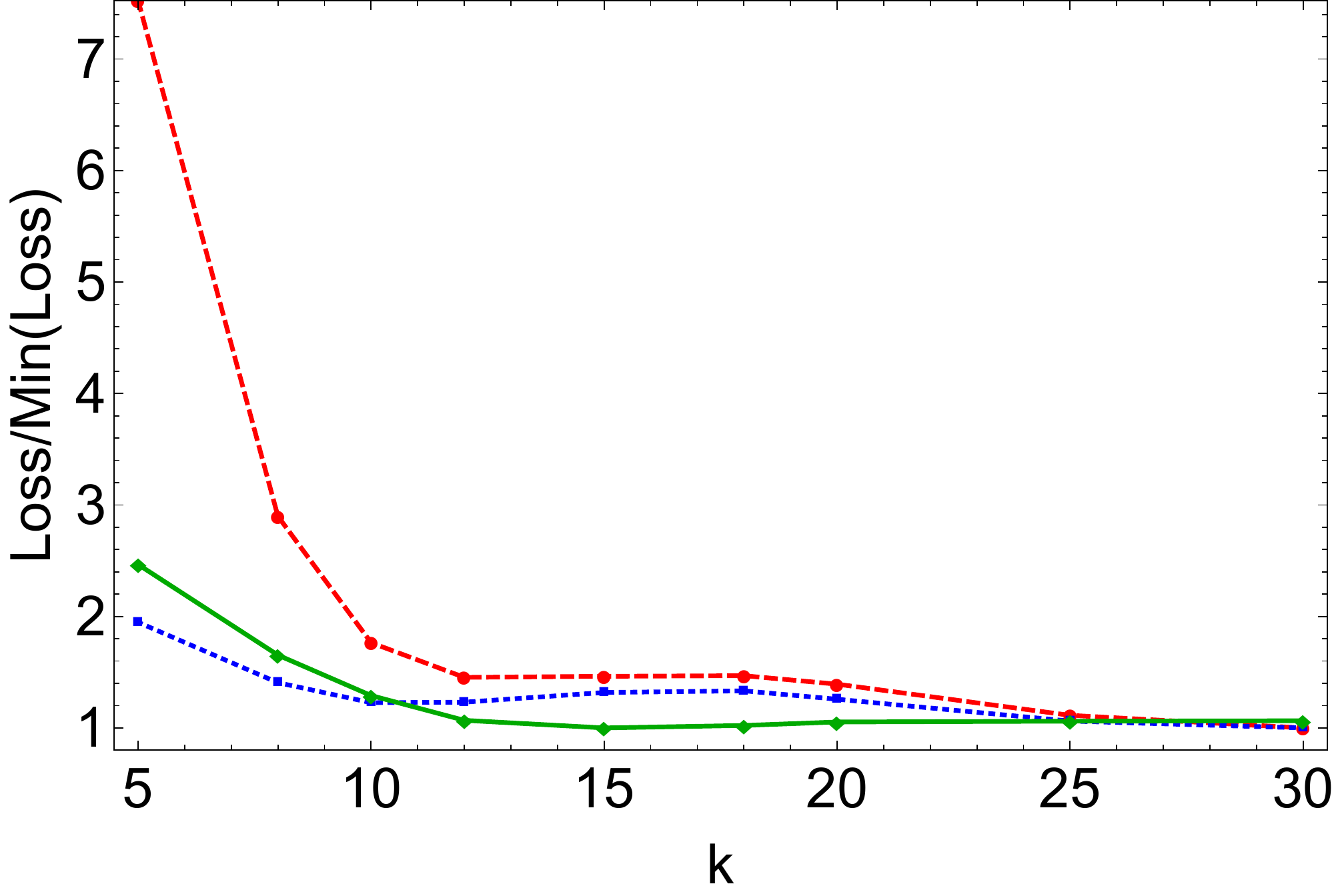}
  \end{center}
  \caption{Same as Fig.~\ref{fig:loss_toy} but for the cosmological model. The 
    number of simulations used here is $d=1000$. Unlike the toy model, there
    isn't a clear optimal banding here, with small changes in the loss for
    banding $k > 10$. 
  }
  \label{fig:loss_cosmo}
\end{figure}

\section{Discussion}
\label{sec:discussion}

This paper describes a method recently introduced in the statistics
literature to directly estimate the precision matrix from an 
ensemble of samples for the case where we have some information about
the sparsity structure of this matrix. This allows for getting
higher fidelity estimates of the precision matrix with relatively 
small numbers of samples.

The {\bf key} result in this paper is the description of an
algorithm to directly estimate the precision matrix without going through
the intermediate step of computing a sample covariance matrix and 
then inverting it. It is worth emphasizing that this estimator is completely
general and does not rely on the sparsity of the precision matrix. 
The estimator does allow us to exploit the structure of the precision matrix
directly; we then use this property for the specific case where the 
precision matrix is sparse. However, we anticipate that this algorithm 
may be useful even beyond the specific cases we consider here.

We also demonstrate the value of regularizing elementwise estimates of the
precision matrix. Although this is not the first application of such 
techniques to precision (and covariance) matrices, we present a concrete
implementation using smoothing splines, including how regularizing parameters
may be automatically determined from the data. 

We demonstrate our algorithm with a series of numerical experiments. 
The first of these, with a explicitly constructed sparse precision matrix, 
allows us to both demonstrate and calibrate every aspect of our algorithm.
Our second set of experiments uses the covariance/precision matrix for
the galaxy two-point correlation function and highlights some of the
real world issues that one might encounter, including the fact that the
precision matrices may not be exactly sparse.
In all cases our method improves over the naive estimation of the sample
precision matrix by a significant factor (see e.g.~Figs.~\ref{fig:toynormbars}
and \ref{fig:cosmonormbars}).
For almost any measure comparing the estimate to truth we find factors of
several improvement, with estimates based on 100 realizations with our method
outperforming the sample precision matrix from 2000 realizations
(see Figs~\ref{fig:toynormd} and \ref{fig:cosmonormd}).
The errors in our method still scales as $N_{\rm sim}^{-1/2}$,
just as for an estimator based on the sample covariance matrix. 
However, our approach achieves this rate for a smaller number of simulations, 
and with a significantly smaller prefactor.

A key assumption in our results is that the precision matrix may 
be well approximated by a banded, sparse matrix. 
This approximation expresses a trade-off between bias and noise.
Banding yields a biased estimate of the precision matrix, but eliminates
the noise in the estimate.
We present a thresholding
procedure to estimate this banding, and find that the 
width of band increases with increasing sample size, as one would expect.
Our banded approximation is similar in spirit to the tapering 
of the precision matrix proposed by \cite{Paz15}. A hybrid approach
may be possible; we defer this to future investigations.

Realistic precision matrices may combine a dominant, approximately sparse
component with a subdominant, non-sparse component. For instance, 
in the case of the correlation function, the non-sparse component
can arise even in Gaussian models from fluctuations near the scale of the
survey volume, from nonlinear effects, as well from the effect of modes outside the survey volume
\citep{Deraps12,dePutter12,Takada13,Kayo13,Mohammed14}. 
In these cases, we imagine combining our estimate of the dominant
term with an estimate of the non-sparse component (perhaps taken directly from
the sample covariance matrix). The key insight here is that our method may be
used to estimate the dominant term with higher fidelity. We defer a detailed
study of methods for estimating the non-sparse component and combining the
estimates to later work.

The computational requirements for the next generation of surveys
is in large part driven by simulations for estimating covariance and precision
matrices. We present an approach that may significantly reduce the number
of simulations required for classes of precision matrices. The ultimate
solution to this problem will likely involve combining model-agnostic
approaches like the one presented in this work, with improved models
for covariance matrices.

\section*{Acknowledgments}

NP thanks Daniel Eisenstein, Andrew Hearin and 
Uro\v{s} Seljak for discussions on covariance matrices. 
NP and MW thank the Aspen Center for Physics, 
supported by National Science Foundation grant PHY-1066293, where part of this 
work was completed. NP is supported in part by DOE DE-SC0008080. HHZ is 
supported in part by NSF DMS-1507511 and DMS-1209191.

\appendix

\section{Useful Linear Algebra Results}
\label{app:la}

For completeness, we include a few key linear algebra 
results used in this paper. We refer the reader to 
\cite{MatrixCookbook} for a convenient reference.

\subsection{The Inverse of a Partitioned Matrix}

Suppose
\begin{equation}
  \mathbf{A} = \left[ \begin{array}{c|c}
  \mathbf{A}_{11} & \mathbf{A}_{12} \\ \hline
  \mathbf{A}_{21} & \mathbf{A}_{22} \end{array} \right]
\end{equation}
then
\begin{equation}
  \mathbf{A}^{-1} = \left[ \begin{array}{c|c}
  \mathbf{B}_1^{-1} &
 -\mathbf{A}_{11}^{-1}\mathbf{A}_{12}\mathbf{B}_2^{-1} \\ \hline
 -\mathbf{B}_2^{-1}\mathbf{A}_{21}\mathbf{A}_{11}^{-1} &
  \mathbf{B}_2^{-1} \end{array} \right]
\end{equation}
where
\begin{eqnarray}
  \mathbf{B}_1 &\equiv& \mathbf{A}_{11}
 -\mathbf{A}_{12}\mathbf{A}_{22}^{-1}\mathbf{A}_{21} \\
  \mathbf{B}_2 &\equiv& \mathbf{A}_{22}
 -\mathbf{A}_{21}\mathbf{A}_{11}^{-1}\mathbf{A}_{12}
\end{eqnarray}

\subsection{Conditional Distributions of a Multivariate Gaussian}

If $\mathbf{x}\sim\mathcal{N}(\mathbf{\mu},\mathbf{C})$ with
\begin{eqnarray}
  \mathbf{x} &=& \left[
  \begin{array}{c} \mathbf{x}_a\\ \mathbf{x}_b\end{array}\right] \\
  \mathbf{\mu} &=& \left[
  \begin{array}{c} \mathbf{\mu}_a\\ \mathbf{\mu}_b\end{array}\right] \\
  \mathbf{C} &=& \left[
  \begin{array}{cc} \mathbf{C}_a   & \mathbf{C}_c \\
                    \mathbf{C}^T_c & \mathbf{C}_b \end{array}\right]
\end{eqnarray}
then $p(\mathbf{x}_a|\mathbf{x}_b)=\mathcal{N}(\mathbf{\mu}'_a,\mathbf{C}'_a)$
with
\begin{eqnarray}
  \mathbf{\mu}'_a &=& \mathbf{\mu}_a + \mathbf{C_cC_b^{-1}}
  (\mathbf{x}_b-\mathbf{\mu}_b) \\
  \mathbf{C}'_a &=& \mathbf{C}_a - \mathbf{C}_c\mathbf{C}_b^{-1}\mathbf{C}^T_c
\end{eqnarray}
Note that the covariance matrices are the Schur complement of the block
matrices.

\section{An Algorithm for Maximum Likelihood Refinement}
\label{app:mlrefine}

The maximum likelihood problem is equivalent to solving
\begin{equation}
  \Bigg[{\cal R}^{-1} - {\cal D} \cS {\cal D} - 2 ({\cal R} - {\cal R}_0)\Bigg]_{\cal J} = 0 \,.
\end{equation}
A variant of this problem was considered in \cite{BakWoerd95}, where they consider 
the solution of above problem without the prior constraint. They suggest using a 
Newton-Raphson algorithm to solve this; we reproduce this below, including the
changes needed for our modified problem.

If ${\cal J} = { (i_1,j_1),\ldots,(i_s,j_s)}$ are the nonzero indices in the ${\cal R}$, we 
define the vectors 
\begin{align}
  x &= \Big( {\cal R}_{i_1 j_1}, \ldots, {\cal R}_{i_s j_s} \Big) \\
  y &= \Big( y_1, \ldots, y_s \Big) \\
  y_p &= \Bigg[{\cal R}^{-1} - {\cal D} \cS {\cal D} - 2 ({\cal R} - {\cal R}_0)\Bigg]_{i_p j_p} 
\end{align}
where $x$ is just the list of elements of ${\cal R}$ we are varying, and $y$ are the 
residuals from our desired solution. The Hessian $H$ is an $s$-by-$s$ matrix with 
elements
\begin{equation}
  H_{pq} = ({\cal R}^{-1})_{i_p j_q} ({\cal R}^{-1})_{i_q j_p} + ({\cal R}^{-1})_{i_q i_p}
    ({\cal R}^{-1})_{j_q j_p} + 2\,.
\end{equation}
The minimization proceeds by iterating the following steps until the residual $||y||_{\infty}$
has reached a pre-determined tolerance (we use $10^{-9}$) :
\begin{enumerate}
  \item Compute the minimization direction $v$ by solving $y = H v$. 
  \item Compute $\delta=\sqrt{v^{T} y}$, and set the step size $\alpha=1$ if $\delta < 1/4$;
    otherwise $\alpha=1/(1+\delta)$.
  \item Update $x \rightarrow x + \alpha v$, and use this to update $\cal R$ and $y$. If the 
    update yields a $\cal R$ that is not positive definite, we backtrack along this direction, 
    reducing $\alpha$ by a factor of 2 each time, until positive definiteness is restored. 
    In practice, this happens rarely, and early in the iteration, and a single backtrack step 
    restores positive definiteness.
\end{enumerate}

The above algorithm is very efficient, converging in $\sim 50$ iterations or fewer for the cases
we consider. However, a major computational cost is in inverting the Hessian\footnote{Even though
we don't explicitly compute the inverse, there is a substantial cost to solving the linear system}.
For a $k$-banded matrix, $s = n(k-1) - k(k-1)/2$; for some of the cases we consider, this is easily 
between $10^{3}$ and $10^{4}$ elements. Inverting the Hessian every iteration is computationally 
too expensive, and we transition over to the {\it BFGS} algorithm \citep{Nocedal2000} for problem
sizes $s > 1000$. Conceptually, the BFGS algorithm (and others of its class) use first derivative
information to approximate the Hessian and its inverse. At each step of the iteration, the inverse 
Hessian is updated using a rank-2 update (Eq. 6.17 in \citealt{Nocedal2000}), based on the change in 
the parameters and the gradient. Since the algorithm works by directly updating the inverse Hessian, 
we avoid the computational cost when computing the minimization direction.

We implement Algorithm 6.1 of \cite{Nocedal2000} and refer the reader
there for a complete description. We limit ourselves here to highlighting the particular choices 
we make. The first is the starting approximation to the Hessian; we use the identity matrix for 
simplicity. While we may get better performance with an improved guess, this choice does not 
appear to adversely affect convergence and we adopt it for simplicity. The second choice is 
how to choose the step size in the one-dimensional line minimization. Motivated by our success 
in the Newton-Raphson case, we use the same step-size choice, with the additional condition that 
we also backtrack if the error on the previous iteration is $< 0.9 \times$ the current error. This 
last condition prevents overshooting the minimum. Note that the {\it Wolfe} conditions 
\citep{Nocedal2000} are automatically satisfied with this algorithm due to the convexity
of our function. 

Finally, we verify that both algorithms converge to the same answer for the
same problems.

\section{Cubic Smoothing Splines and Cross Validation}
\label{app:cubic}

We summarize the construction of cubic smoothing splines as used here; our
treatment here follows \cite{Craven1979} (see also \citealt{Reinsch1967, DeBoor2001}).

Consider a function $y_i \equiv y(i)$ evaluated at $n$ evenly spaced\footnote{This condition
is not necessary, but is what is relevant for our application} points $i=1,\ldots,n$. We 
aim to find a function $f(x)$ that minimizes 
\begin{align}
  p \sum_{i=1}^{n} (f(i) - y_i)^2 + \int_{1}^{n} dx \,(f''(x))^2
\end{align}
where $p$ balances fidelity to the observed points and the smoothness of the function, 
and is an input parameter. For appropriate conditions\footnote{Square integrability, and
continuous second derivatives}, the solution to this is a cubic spline, with points 
at $i=1,\ldots,n$. Our goal therefore reduces to determining $p$ and $f_i \equiv f(i)$.

\cite{Craven1979} suggest a variation on cross-validation to determine the value
of $p$. In ordinary cross-validation, one removes a point at a time from the 
data and minimizes the squared deviation (over all points) of the spline prediction for the dropped 
point from its actual value. While conceptually straightforward, actually performing
this cross-validation is operationally cumbersome. \cite{Craven1979} suggest a weighted 
variant of ordinary cross-validation (generalized cross-validation), that both 
theoretically and experimentally, has very similar behaviour, and is straightforward to compute.
We outline this method below.

The $f_i$ are determined from the $y_i$ by 
\begin{align}
\begin{pmatrix} f_1 \\ .\\ .\\ f_n \end{pmatrix} 
 &= {\bf A}(p) 
\begin{pmatrix} y_1 \\ .\\ .\\ y_n \end{pmatrix} 
\end{align}
where ${\bf A}(p)$ is an $n \times n$ matrix that depends on $p$. Following
\cite{Reinsch1967} and \cite{Craven1979}, we construct ${\bf A}(p)$ as follows :
\begin{enumerate}
\item Construct the $n \times n-2$ tridiagonal matrix ${\bf Q}$ with the 
  following non-zero elements $q_{i,i+1} = q_{i+1,1} = 1$ and $q_{ii} = -2$,
  where $i = 1, \ldots, n$. 
\item Construct the $(n-2) \times (n-2)$ tridiagonal matrix ${\bf T}$ with 
  non-zero elements $t_{n-2,n-2} = t_{ii} = 4/3$ and $t_{i,i+1} = t_{i+1,i} = 2/3$, with
  $i=1,\ldots,n-3$.
\item Compute ${\bf F} = {\bf Q} {\bf T}^{-1/2}$. Construct its singular value decomposition ${\bf F} = {\bf U} {\bf D} {\bf V}^{t}$
  with ${\bf D}$ a diagonal matrix with $n-2$ singular values $d_i$, and ${\bf U}$ and $\bf V$ being
  $n \times (n-2)$ and $(n-2) \times (n-2)$ orthogonal matrices respectively.
\item Define the $n-2$ values $z_j$ by ${\bf U}^{t} {\bf y}$, where ${\bf y}$ are the $y_i$ arranged in a column
  vector as above. Define $\tilde{d}_i \equiv d_i^2/(d_i^2+p)$. 
\item The generalized cross-validation function $V(p)$ is given by
  \begin{equation}
    V(p) = \frac{1}{n} \sum_{i=1}^{n-2} \tilde{d}_i^2 z_i^2 \Bigg/ \Bigg( \sum_{i=1}^{n-2} \tilde{d}_i \Bigg)^2 \,.
  \end{equation}
  We minimize this function using a simple linear search in $\log p$; the minimum determines the value of $p$.
\item The cubic spline matrix is then given by ${\bf I} - {\bf A} = {\bf U} {\bf \tilde{D}} {\bf U}^{t}$, where
${\bf \tilde{D}}$ is an $(n-2) \times (n-2)$ diagonal matrix with $\tilde{d}_i$ on the diagonal.
\end{enumerate}

\bibliographystyle{mn2e}
\bibliography{sparse1}

\end{document}